\documentclass[twocolumn]{aastex62}

\usepackage{graphicx}
\usepackage{natbib}
\usepackage{multirow}
\usepackage{textcomp}
\usepackage{longtable}
\usepackage{tabularx}
\usepackage{ulem}
\usepackage{soul} 
\usepackage{enumitem}
\usepackage{chngcntr}
\usepackage{hyperref}
\usepackage{changepage}
\usepackage{amsmath}
\hypersetup{
    colorlinks=true,
    filecolor=magenta,      
    urlcolor=blue,
}

\newcommand{\bc}{\begin{center}}
\newcommand{\ec}{\end{center}}

\newcommand{\degree}{$^{\circ}$}
\newcommand{\msun}{M$_{\odot}$}

\newcommand{\ra}{$\alpha$(2000)}
\newcommand{\dec}{$\delta$(2000)}
\newcommand{\h}{$^{\mathrm{h}}$}
\newcommand{\m}{$^{\mathrm{m}}$}
\newcommand{\s}{$^{\mathrm{s}}$}

\newcommand{\kms}{km s$^{-1}$}
\newcommand{\kmsp}{km s$^{-1}$ pc$^{-1}$}

\newcommand{\jyb}{Jy b$^{-1}$}
\newcommand{\jybe}{Jy beam$^{-1}$}

\newcommand{\am}{NH$_{3}$}

\newcommand{\cyano}{HC$_3$N}
\newcommand{\meth}{CH$_3$OH}

\newcommand{\methcy}{CH$_3$CN}

\newcommand{\cyanobut}{HC$_5$N}

\newcommand{\mtrans}{$4_{\textrm{-}1}$$-$$3_{0}$}

\newcommand{\scloud}{M0.20$-$0.033}

\newcommand{\rar}{the Radio Arc region} 

\newcommand{\mshell}{\scloud~expanding shell}
\newcommand{\stone}{M0.07$-$0.07}
\newcommand{\sticks}{M0.10$-$0.08}

\newcommand{\gecloud}{M0.11$-$0.11}

\newcommand{\fifty}{M$-$0.02$-$0.07}
\newcommand{\sgra}{Sgr A$^*$}

\newcommand{\pv}{position-velocity}

\newcommand{\mc}{molecular cloud}

\newcommand{\gc}{Galactic center}

\newcommand{\clumpfind}{\textit{Clumpfind}}

\newcommand{\xray}{X-ray}

\newcommand{\ie}{i.e.}
\newcommand{\eg}{e.g.}
\newcommand{\til}{$\sim$}
\newcommand{\andd}{\&}
\newcommand{\leftit}{\textit{left}}

\begin{document}

\title{\uppercase{Evidence for an interaction between the Galactic Center clouds: \sticks~and \gecloud}} 

\author[0000-0002-4013-6469]{Natalie O. Butterfield}
\affil{Department of Physics, Villanova University, 800 E. Lancaster Ave., Villanova, PA 19085, USA}
\affil{National Radio Astronomy Observatory, 520 Edgemont Road, Charlottesville, VA 22903, USA}
\email{nbutterf@nrao.edu}

\author{Cornelia C. Lang}
\affil{Department of Physics and Astronomy, University of Iowa, 30 North Dubuque Street, Iowa City, IA 52242, USA}

\author[0000-0001-6431-9633]{Adam Ginsburg}
\affil{Department of Astronomy, University of Florida, PO Box 112055, USA}


\author[0000-0002-6753-2066]{Mark R. Morris}
\affil{Department of Physics \& Astronomy, University of California, 430 Portola Plaza, Los Angeles, CA 90095, USA} 

\author[0000-0001-8224-1956]{J\"{u}rgen Ott}
\affil{National Radio Astronomy Observatory, 1003 Lopezville Road, Socorro, NM 87801, USA}

\author[0000-0002-8631-6457]{Dominic A. Ludovici}
\affil{Department of Physics and Engineering Science, Coastal Carolina University, 100 Chanticleer Dr E, Conway, SC 29528, USA}

\begin{abstract}

We present high-resolution (\til2$-$3\arcsec; \til0.1 pc) radio observations of the Galactic center cloud \sticks~using the Very Large Array at K and Ka band (\til25 and 36 GHz). The \sticks~cloud is located in a complex environment near the Galactic center Radio Arc and the adjacent \gecloud~molecular cloud. From our data, \sticks~appears to be a compact molecular cloud (\til 3 pc) that contains multiple compact molecular cores (5+; $<$0.4 pc). In this study we detect a total of 15 molecular transitions in \sticks~from the following molecules: NH$_3$, HC$_3$N, CH$_3$OH, HC$_5$N, CH$_3$CN, and OCS. We have identified more than sixty 36 GHz \meth~masers in \sticks~with brightness temperatures above 400 K and 31 maser candidates with temperatures between 100$-$400 K. We conduct a kinematic analysis of the gas using \am~and detect multiple velocity components towards this region of the Galactic center. The bulk of the gas in this region has a velocity of 51.5 \kms~(\sticks) with a lower velocity wing at 37.6 \kms. We also detect a relatively faint velocity component at 10.6 \kms~that we attribute {to being} an extension of the \gecloud~cloud. Analysis of the gas kinematics, combined with past X-ray fluorescence observations, {suggests} \sticks~and \gecloud~are located in the same vicinity of the Galactic center and could be physically interacting.

\end{abstract}


\section{Introduction}

\begin{figure*}[tb!]
\centering
\includegraphics[scale=0.45]{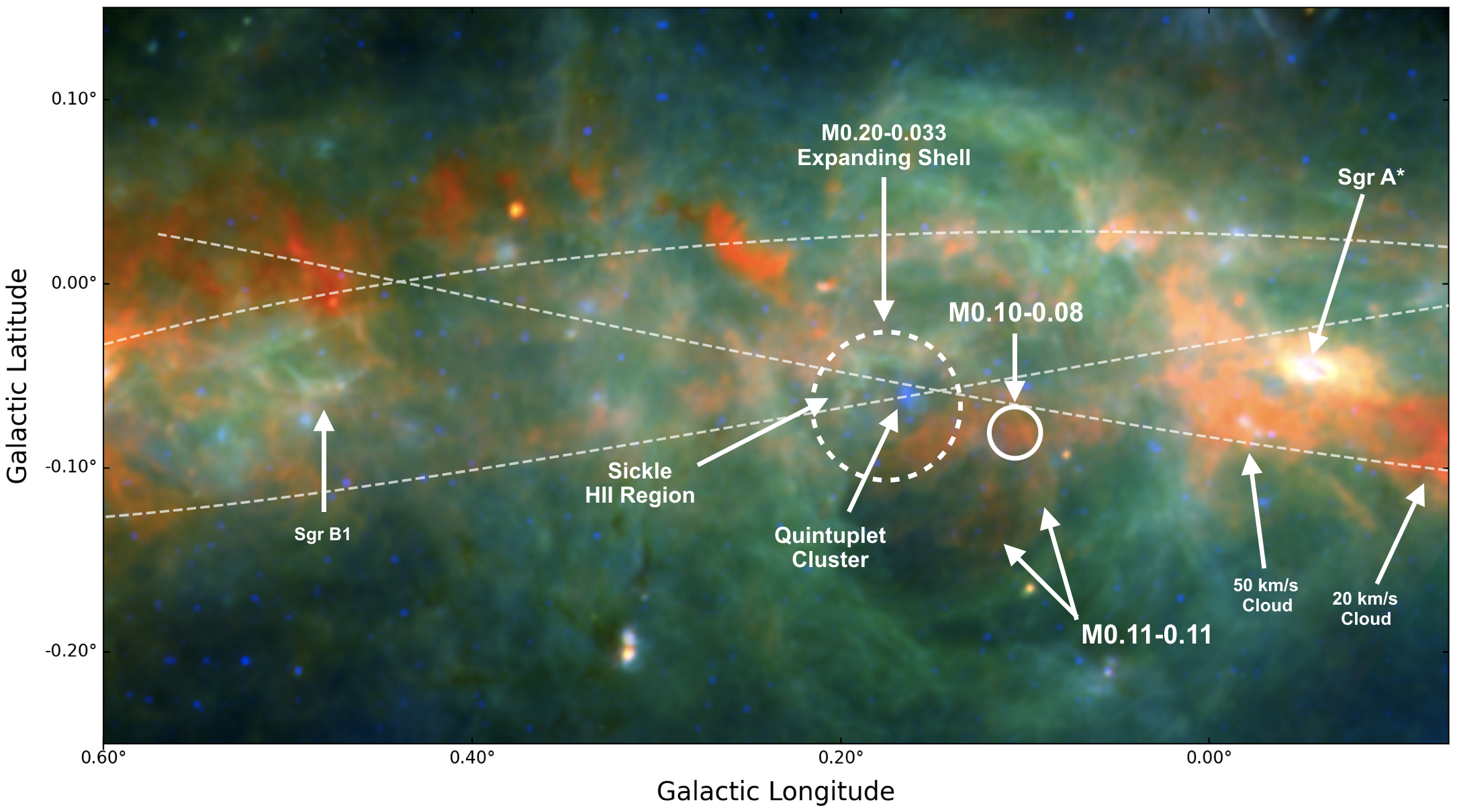}  
 \caption{ 
{Three-color} composite of the inner 100 pc of the CMZ, centered on \rar, {where red and green are the 160 and 70 $\mu$m emission, respectively, from HiGAL \citep{Molinari10}}, and blue is 8 $\mu$m emission from Spitzer {\citep{Churchwell09}}. The solid white circle shows the region of the CMZ targeted in this study. This field is centered on the \sticks~molecular cloud, but could overlap with some of the extended emission in \gecloud.  The dashed white circle shows the location of the \mshell\ {presented in \cite{my17}}. Additional prominent CMZ regions are labeled for reference purposes. {Overlaid on this figure is a dashed line showing the extent of the orbital stream proposed by \cite{Kru15}.} }
\label{introfig}
\end{figure*}

The central 200 pc of the galaxy (Central Molecular Zone; CMZ) is an extreme Galactic environment. Molecular clouds in the CMZ have hotter average gas temperatures \citep[50$-$300 K;][]{Mauers86, mills13, Krieger17, Ginsburg16}, higher densities \citep[10$^{3-5}$ cm$^{-3}$;][]{Zylka92, mills18a}, and broader line widths, on the \til10 pc scale \citep[\til20$-$30 \kms;][]{bally87, Kauffmann17a}, than typical clouds in the interstellar medium (ISM) of the Galactic disk. The velocities of CMZ clouds range from $-$250 to +250 \kms~within the inner 1\fdg5 of our Galactic center. The large velocity range of these clouds, wide velocity dispersions, and line-of-sight confusion from multiple velocity components can make it difficult to place individual molecular clouds within the 3-dimensional context of the CMZ.   
{Figure \ref{introfig} shows the inner 100 pc of the Galactic center, where many of these dense molecular clouds are shown in red in this 3-color image. }
Recent efforts have been made to connect {these} individual clouds (1$-$10 pc) to the larger structures (\til100 pc) in the Galactic center {\citep[][]{Sofue95, sawada04,Molinari11, Kru15,Henshaw16}.}

{The {3-dimensional} orientation of the large-scale structures in the CMZ can depend greatly on the interpretation of the gas kinematics. For example, \cite{Sofue95} and \cite{sawada04} suggest a two spiral arm structure, whereas \cite{Molinari11} argue for a twisted elliptical ring. The most recent orbital model, presented in \cite{Kru15}, suggests an open orbit solution (see dashed line in Figure \ref{introfig} for the projected trajectory of their orbital model solution).} 
In the {\cite{Kru15} orbital} model, gas in the CMZ traces an open orbit set by the shape of the CMZ potential. Connected chains of molecular clouds all follow the same orbital path or `stream'.  The {3-dimensional} arrangement of clouds along a continuous stream can be loosely reconstructed from their projected radial distance to \sgra~and the observed line-of-sight velocity.
However, there is still some ambiguity about whether certain features are located on the near or far sides of the Galactic center. Additionally, multiple components along the same line-of-sight can make it challenging to disentangle the kinematics of a single cloud.
High spatial and spectral resolution observations targeting regions where the kinematics are complex are needed to resolve the {individual} components.

\begin{table*}[bt!]
\caption{\textbf{Spectral Line Imaging Parameters for the 15 molecular transitions detected in this study}}
\centering
\begin{tabular*}{\textwidth}{l @{\extracolsep{\fill}} ccccccccc}
\hline\hline
 & & \multicolumn{3}{c}{---Restoring Beam\footnote{\label{note1} Where 1\arcsec~is 0.04 pc at an assumed distance of 7.9 kpc to the CMZ \citep{Do19}.}---} & &     \\
 Species \& & Rest  & ~Major~  & ~Minor~  & ~Position~  & Velocity & ~{rms}~ & Peak \\
  Transition & ~Frequency~  & Axis & Axis & Angle  & ~Resolution~ & {per Channel} & Intensity \\
 & (GHz) & (\arcsec) & (\arcsec)  & (\degr) &  (\kms) & ~(m\jybe)~ & (m\jybe)  \\
\hline
\am~(1,1) & 23.69450  & 2.81 & 2.62 & -81.85 &  1.58 & 1.3 & 57.6   \\
\am~(2,2) & 23.72263 & 2.79 & 2.62 & -85.45  &  1.58 & 1.4 & 56.3 \\
\am~(3,3) &  23.87013  & 2.77 & 2.60  & -85.18   & 3.14 & 2.3 & 485.9   \\
\cyanobut~(9$-$8)\footnote{\label{note2} These four transitions were smoothed from the natural spatial resolution to improve the signal-to-noise.}\footnote{\label{note3} These four transitions had a signal-to-noise ratio $<$9 and are therefore not shown in Figure \ref{morphfig}.}  &        23.96390    & 3.00  & 3.00  & 0.00  & 3.13 & 0.8 & 6.4 \\
\am~(4,4)   & 24.13942 & 2.79 & 2.59  & 89.49  & 3.10 & 1.0  & 42.0  \\
OCS (2$-$1)\textsuperscript{\ref{note2}}\textsuperscript{\ref{note3}}   &       24.32593    & 5.00  & 5.00 & 0.00 & 3.08 & 1.2 &  8.6 \\ 
\am~(5,5)  & 24.53299  & 2.74 & 2.55  & -90.00  &    3.05 & 0.8  & 28.3 \\
\meth~(6$_2-$6$_1$)\textsuperscript{\ref{note2}}\textsuperscript{\ref{note3}}~~~~~~~ & 25.01812  & 3.00  & 3.00  &  0.00 &  3.00 & 0.8 & 5.6   \\
\am~(6,6) &  25.05603    & 2.63  & 2.48 & 89.48 &    2.99 & 0.9  & 29.4  \\
\am~(7,7)  & 25.71518    & 2.58 & 2.40 & -83.33  & 2.91 & 0.7 & 11.8  \\ 
\cyano~(3$-$2) & 27.29429  & 3.14  & 2.52  & 16.76  &    2.75  & 2.2  & 55.2  \\
\am~(9,9)\textsuperscript{\ref{note3}}  &   27.47794  &  2.80 & 2.45  & 43.30  & 2.73  & 2.0 & 16.8  \\
\meth~(\mtrans) & 36.16927 &   1.95  & 1.81 & -174.63  &   1.05  & 2.9 (50)\footnote{The larger value in parentheses is the rms noise in the channel containing the brightest maser, at v = 50.6 \kms.} &  46969.0      \\
\cyano~(4$-$3)  &  36.39232 &  1.98 & 1.81  & -3.02  & 2.06  & 2.8 & 47.0  \\
\methcy~(2$-$1)\textsuperscript{\ref{note2}} &    36.79547     & 3.00  & 3.00  & 0.00 & 1.02 & 4.3 & 40.7 \\
\hline
\end{tabular*}
\label{Images}
\end{table*}




One region of the CMZ where the kinematics are complex is toward the \sticks~\mc\ {(solid white circle in Figure \ref{introfig})}. 
The \sticks~cloud, and the adjacent \gecloud\ cloud (annotated in Figure \ref{introfig}), have been observed in several large-scale surveys of molecular gas in the CMZ for many decades {\citep{Gusten81, tsuboi97, chuss03, Handa06, Jones12, mills17, cmzoom1, cmzoom2, guan21}. }
{Several of the} low-spatial-resolution surveys of cold dust and molecular gas show that the \sticks~cloud is relatively bright and compact ($<$3 pc), {with a mass of 1.7 $\times$ 10$^5$ \msun~\citep[{\eg,}][]{tsuboi11}}. {The \sticks~cloud also been observed to have substructure, as detected in the recent 1mm CMZoom survey \citep{cmzoom1, cmzoom2}}. \gecloud, however, is relatively faint and extended  ($>$5 pc) and could spatially overlap with \sticks~(Figure \ref{introfig}). 
The spatial overlap between the two clouds has led some {investigators} to argue for a possible connection between the two clouds \citep{Handa06, clavel13}. However, there are {unsolved questions} about this connection in the literature due to the large velocity separation between the two clouds {along this line-of-sight \citep[$\Delta$v\til30 \kms;][]{ponti10, Kru15}.} 
{Understanding the connection or separation of the two clouds can give insight on the complex kinematics in the region. Furthermore, disentangling the complex kinematics into a somewhat simple solution is essential for understanding the {3-dimensional} structure of the gas and the effects that cloud-cloud interactions can have on the gas motions.}

We present high-resolution (\til2$-$3\arcsec) radio observations of \sticks~using the National Science Foundation's Karl G. Jansky Very Large Array (hereafter, VLA). Using these observations we analyze the morphological and kinematic structure of \sticks~at high-resolution (Section \ref{results}) and discuss the relationship of \sticks~to other clouds in the region (Section \ref{dis}).


\section{Observations and Data Calibration} 
\label{obs}

\begin{figure*}[tb!]
\centering
\includegraphics[scale=0.82]{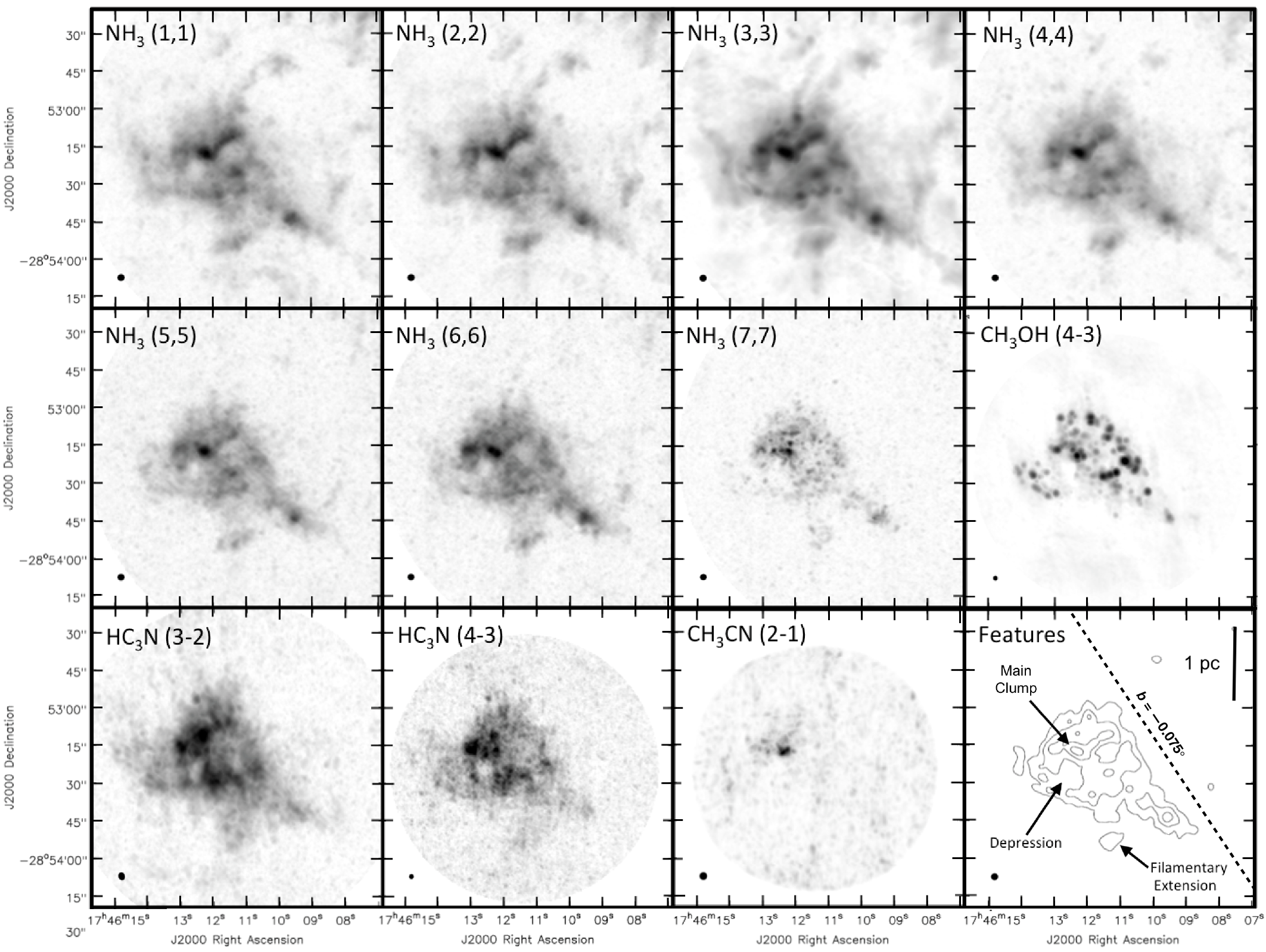} 
\caption{{Peak} intensity distribution of 11 of the 15 molecular line transitions detected in this paper.  The top two rows show the \am~(1,1) $-$ (7,7) and \meth~{line emission}. The bottom row shows the observed \cyano~and \methcy~transitions. The bottom right-most panel shows the 20, 40, 80, and 140 $\sigma$ contour levels of the \am~(3,3) {emission}, with {annotations identifying} several {of the} `Features' discussed in Section \ref{g10morph}.  The spatial resolution of each presented molecular transition is shown in the bottom left corner of every panel. The imaging parameters of all 15 detected molecular transitions are described in Table \ref{Images}. The black dashed line shows the orientation of the Galactic plane at $b=-0\fdg075$. }
\label{morphfig}
\end{figure*}

The observations presented in this paper were taken with the VLA interferometric radio telescope, operated by the National Radio Astronomy Observatory.\footnote{The National Radio Astronomy Observatory is a facility of the National Science Foundation operated under cooperative agreement by Associated Universities, Inc.} These VLA observations were part of a larger survey of \mc s in the CMZ (PI: Elisabeth A.C. Mills; Project code: 11B-210).\footnote{Results from this survey have also been presented in: \cite{mills13, mills14, mills15, dom16, my17, mills18b}.} This survey used the K (18.0$-$26.5 GHz) and~Ka (26.5$-$40.0 GHz) band receivers on 2012 January 14$^{\mathrm{th}}$ \& 13$^{\mathrm{th}}$, respectively, with the DnC hybrid array. In this survey we observed 15 spectral lines {from} several regions in the CMZ. The image cube parameters for all 15 lines are reported in Table \ref{Images}. The results presented in this paper focus on a single pointing containing \sticks,\footnote{All J$<$7 \am~images, shown in Figure \ref{morphfig}, are from a larger multi-pointing mosaic \citep[See Figure 3, left, in][]{my17}.} centered at \ra=17$^{\mathrm{h}}$46$^{\mathrm{m}}$09\fs79, \dec=$-28\degr 53\arcmin 18\farcs0$, for K band, and \ra=17$^{\mathrm{h}}$46$^{\mathrm{m}}$11\fs37, \dec=$-28\degr 53\arcmin 24\farcs3$, for Ka band, with a time-on-source of \til25 minutes in each frequency band.

The correlator setup for this survey is described in \cite{mills15} and \cite{my17}. High-frequency VLA procedures\footnote{\href{https://casaguides.nrao.edu/index.php?title=EVLA_high_frequency_Spectral_Line_tutorial_-_IRC\%2B10216_part1}{Hyperlink to the high frequency CASA tutorial}. All imaging and calibration of the VLA observations presented here used the Common Astronomy Software Application (CASA) program provided by NRAO \citep{Casa}.} were used for calibration and imaging, as described in \cite{mills15}, with one difference. We employed the CLEAN parameter ``multiscale'' for all spectral lines that had a signal-to-noise ratio $>$15 and a peak intensity $>$20 m\jybe~(see Table \ref{Images}) in order to improve our sensitivity to {data taken with} short baselines in our interferometric {observations}.


\section{Results}
\label{results}

\subsection{Morphology of the Molecular Emission in \sticks}
\label{mol-res}

Figure \ref{morphfig} presents the {peak} intensity emission of 11 molecular transitions detected in M0.10$-$0.08 (see Table \ref{Images} for imaging parameters). The remaining four detected molecular transitions in \sticks~are {relatively} faint ($<$9$\sigma$) and are therefore $not$ shown in Figure \ref{morphfig}. In the following sections we examine the bright, diffuse molecular emission (\S \ref{g10morph}: \am~\& \cyano). We focus on the kinematics of the \am~emission and fit {the averaged gas profile} in Section \ref{g10kin}. The \meth~(\mtrans) class I maser transition is discussed in detail in Section \ref{masertext}.


\subsubsection{Morphology of the Diffuse Molecular Emission: \am~and \cyano}
\label{g10morph}

The top two rows in Figure \ref{morphfig} show the detected NH$_3$ (1,1)$-$(7,7) emission in M0.10$-$0.08. The distribution of the metastable \am~emission is similar across all seven~transitions. The speckled morphology observed in the \am~(7,7) transition is likely an artifact of cleaning {with delta functions (see Section \ref{obs} and Table \ref{Images} for a discussion on the cleaning process)}.

Most of the \am~emission in the \sticks~cloud~is concentrated within a square arcminute region near the center of the field.  At high-resolution (3\arcsec) the \sticks~cloud has a wedge-like appearance that is narrow at lower Galactic longitude (10\arcsec; $l$=0\fdg095) and widens with increasing Galactic longitude (50\arcsec; $l$=0\fdg11). This wedge-like structure is also noticeable in both transitions of \cyano: 3$-$2 and 4$-$3 (bottom row in Figure \ref{morphfig}). Additionally, there is a diffuse `filamentary extension' toward the southern region of \sticks,~as indicated in the bottom right-most panel of Figure \ref{morphfig} (\ie, `Features' panel). This filamentary extension is detected in both the \am~and \cyano~transitions, but not in the \meth~(4$-$3) transition. The longest extent of \sticks~is 75\arcsec~(\til3 pc), indicating that this cloud~is among the more compact of molecular clouds observed in the \gc~{\citep[diameters of 3$-$10 pc; \eg,][]{Gusten81,bally87,Kauffmann17a,mills17}}.

\begin{figure}
\includegraphics[scale=0.2]{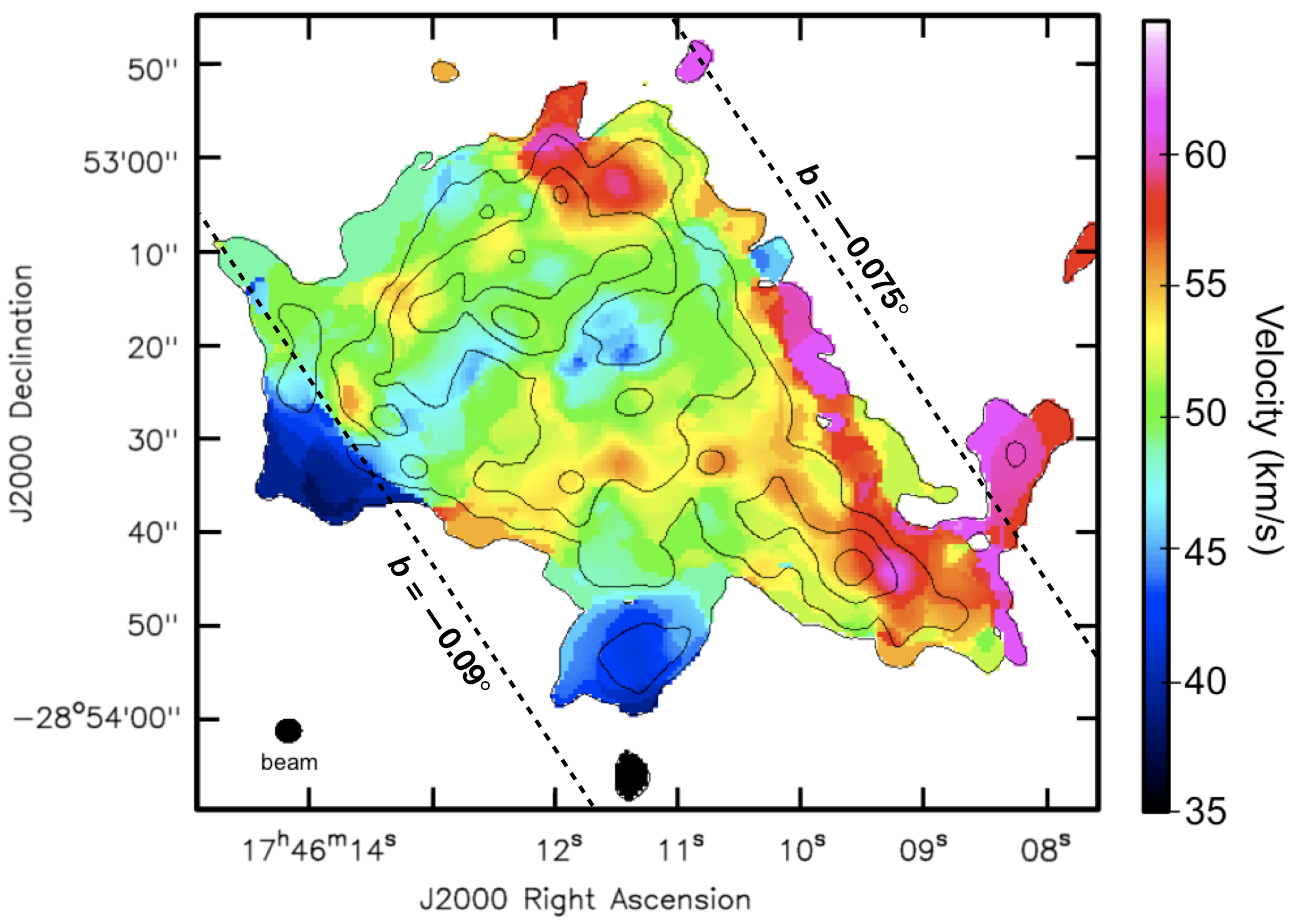} 
\caption{
{Intensity-weighted} velocity distribution (1st moment map) of the \am~(3,3) transition for emission above the 10$\sigma$ level, integrated over a velocity range of {-20}$-$100 \kms. The black contours correspond to emission at 10, 20, 40, 80, \andd~140 $\times$ 2.3 m\jybe~(rms level). The black {dashed lines show} the orientation of the Galactic plane at $b=-0\fdg075$ {and $b=-0\fdg09$}.}
\label{spectrum}
\end{figure}

Within \sticks~there are several (\til5) compact {clumps (D$<$10\arcsec)} of brighter \am~emission {($>$0.2 \jybe)} that are most prominent in the (3,3) transition. Most of these compact clumps are concentrated toward the northeast region of the cloud, with the brightest \am~clump located at: \ra=17\h46\m12\fs3,~\dec= $-$28\degr53\arcmin18\arcsec. The brightest clump (\ie, `Main Clump'; see `Features' panel in Figure \ref{morphfig}) contains emission in all 11 transitions shown in Figure \ref{morphfig}, including the fainter \methcy~(2$-$1) transition. Further, the Main Clump is the only location where we detect \methcy~emission. Directly south of the Main Clump is a lower intensity emission region (`Depression;' labeled in the Features panel of Figure \ref{morphfig}).  This depression region is \til10\arcsec~across and is located at \ra=17\h46\m12\fs5, \dec=$-$28\degr53\arcmin25\arcsec. The Depression is detected in both \am~and \cyano, but is most {prominent} in the \cyano~(4$-$3) transition.\footnote{There is a second lower level emission region to the west of the Main Clump and north of the Filamentary Extension (see Figure \ref{morphfig}). However, because we detect emission above the noise level in this region in the \cyano~(4$-$3) transition, we do not characterize this feature as a second `depression'.} Although this feature is detected in all of our extended emission lines (\am, \cyano), it could be produced by spatial filtering in our interferometer data. Future observations at different wavelengths are necessary to determine whether the Depression is some kind of cavity.

\begin{figure}
\centering
\includegraphics[scale=0.57]{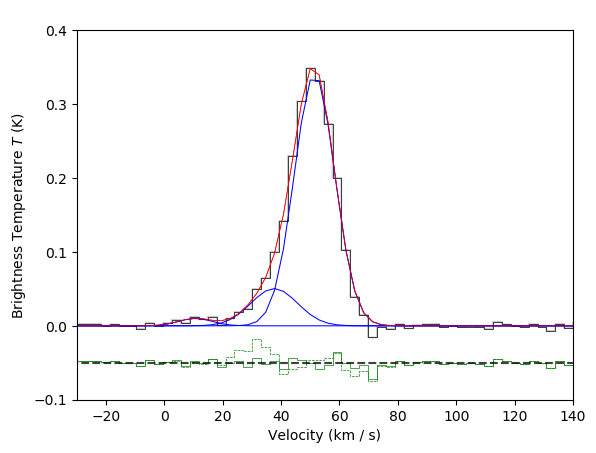} 
\caption{\am~(5,5) velocity spectrum {averaged} over the entire field of view. The \am~(5,5) line was chosen as a representative {spectrum} to show the multiple components toward this region. In the (5,5) line all three components are detected and at this higher J transition, the hyperfine lines are suppressed. The black line shows the data. The blue Gaussians show the individual components (presented in Table \ref{55table}), with the red line showing the sum of the three Gaussian components. The solid green line at -0.05 K shows the residuals of the {three-Gaussian-component} fit. The dashed green line at -0.05 K shows the residuals of a {two-Gaussian-component} fit {(v$_c$ $\simeq$ 10 \kms~and 50 \kms)}. In the two Gaussian fit there is consistent excess of emission around \til20$-$40 \kms~(six spectral cube channels; dashed green line). {This excess emission around 20$-$40 \kms~is brighter than the emission in the 10.6 \kms~component and is detected in the \cyano~transitions as well.} Therefore, we {interpret} the excess emission as an intermediate-velocity component.  
}
\label{am-spectra}
\end{figure}



\subsection{Kinematics of the \am~Emission}
\label{g10kin}

Figure \ref{spectrum} shows the centroid velocity distribution (moment 1) of the \am~(3,3) transition. Most of the bright \am~emission ($>$10$\sigma$) is at a velocity of 35$-$65 \kms. However, as we will show in the following section, faint ($<$10$\sigma$) molecular emission is detected {at lower} velocities of \til10 \kms. 

We note an asymmetry in the velocity distribution that results in roughly a 10 \kmsp~gradient (where 1 pc is \til25\arcsec). Most of the higher-velocity \am~(3,3) emission ($v$ $\geq$ 55 \kms) is located toward the {north-western} side of \sticks~({around \textit{b}=$-$0\fdg075}) and the lower-velocity emission ($v$ $\leq$ 45 \kms) is generally located toward the south and {south-eastern} sides of \sticks~({around \textit{b}=$-$0\fdg09}). The orientation of the described velocity gradient is perpendicular to the direction of orbital motion {in the \cite{Kru15} orbital model}. The filamentary extension, described in Section \ref{mol-res}, contains mainly lower velocity emission {(35$-$45 \kms)} and is oriented roughly parallel to the described velocity gradient. 

\begin{table}[bt!]
\caption{\textbf{Kinematics of the \am~(5,5) transition}}
\centering
\begin{tabular}{lcc}
\hline\hline
\textbf{Parameter}\footnote{$v_c$ is the central velocity of the component, $\sigma$ is the velocity dispersion and $T_{B}$ is the peak brightness temperature.} ~ ~ ~ ~~ ~~ ~	 & ~ ~ ~ ~~ ~ ~ ~	&  \textbf{Value}  \\
\hline
\multicolumn{3}{l}{\textbf{Low Velocity Component}} \\ [0.05cm]
\hline
~ ~ ~$v_{c}$  	&  ~ ~ ~ ~~ ~	&  10.6 $\pm$ 2.5 \kms 	  \\
~ ~ ~$\sigma$ 	& 	& 5.8 $\pm$ 2.6 \kms	 \\
~ ~ ~$T_{B}$   	& 	&  0.01 $\pm$ 0.003 K	 \\
\hline
\multicolumn{3}{l}{\textbf{Intermediate Velocity Component }} \\ [0.05cm]
\hline
~ ~ ~$v_{c}$ 	&	&  37.6 $\pm$ 5.7 \kms	   \\
~ ~ ~$\sigma$ 	& 	&  8.0 $\pm$ 3.0 \kms 	 \\
~ ~ ~$T_{B}$  		& & 0.05 $\pm$ 0.02 K   \\
\hline
\multicolumn{3}{l}{\textbf{High Velocity Component}} \\ [0.05cm]
\hline
~ ~ ~$v_{c}$ 	&	&  51.5 $\pm$ 0.6 \kms 	 \\
~ ~ ~$\sigma$ 	& 	&  6.9 $\pm$ 0.3 \kms 	  \\
~ ~ ~$T_{B}$  		& 	& 0.34 $\pm$ 0.03 K	  \\ 
\hline\hline
\end{tabular}
\label{55table}
\end{table}


\subsubsection{Multiple velocity components toward \sticks}
\label{3comp}

Moment 1 maps, like the one presented in Figure \ref{spectrum}, have the advantage of showing the {predominant} velocity distribution across a cloud or region. However, these maps can be misleading since they can average over multiple components and be weighted by the brighter emission components. Integrating the emission across a cloud or region, and analyzing the spectra using fitting programs like \textit{pyspeckit} \citep{2011ascl.soft09001G,ginsburg22}\footnote{The $pyspeckit$ python program is available online at \href{https://github.com/pyspeckit/pyspeckit}{https://github.com/pyspeckit/pyspeckit}. } can help identify and distinguish multiple components. Once these velocity components are disentangled we can map their spatial distribution and morphology by isolating channels associated with the individual velocity component. {By analyzing the gas kinematics using numerous methods we can understand the {relative structure of the two clouds} towards this complex region.} In this section we will identify the velocity components towards the \sticks~cloud by analyzing the \am~(5,5) line.

\begin{figure}
\includegraphics[scale=0.47]{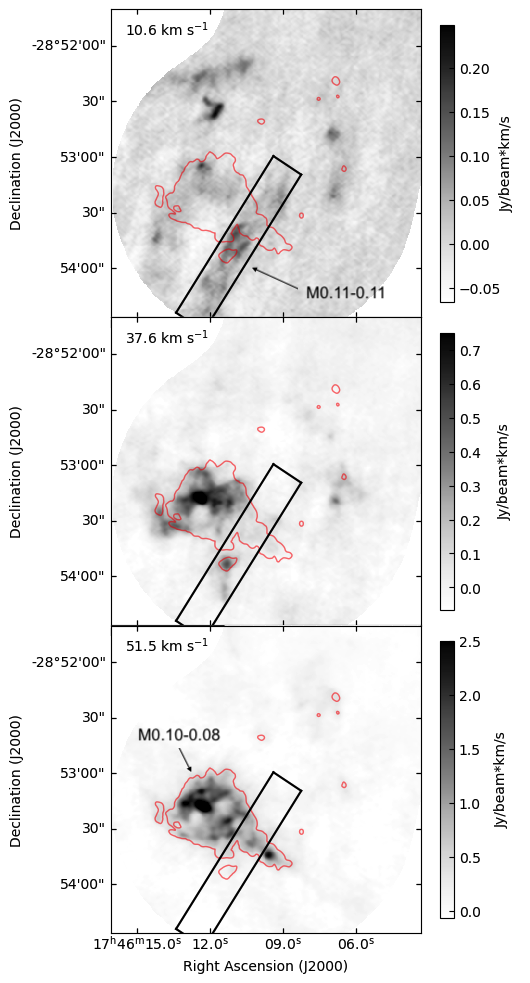} 
\caption{Molecular morphology of the three velocity components presented in Table \ref{55table} (integrated intensity, moment 0, in \am~(3,3)). 
These panels were made using V$_c$ $\pm$ $\sigma_v$ (to the closest channel). The red contour in all three panels shows the 20$\sigma$ level from the `Features' panel in Figure \ref{morphfig} for spatial reference. {Annotated on the 10.6 and the 51.5 \kms\ panels are the \gecloud\ and \sticks\ clouds, respectively}
The black box shows the {region used for the} \pv~slice in Figure \ref{g10pv}.} 
\label{velocity-panels}
\end{figure}

Figure \ref{am-spectra} shows the raw integrated spectrum (black histogram) of the \am~(5,5) line. We chose to analyze the J=5 \am~transition because the hyperfine {satellite} lines are {quite weak} and do not contribute significantly to the spectrum. We initially fit the \am~(5,5) line with two main Gaussian components at \til10$-$15 \kms~and \til50$-$55 \kms. The residuals from this initial fit showed an excess around 20$-$40 \kms~(dashed green line; Figure \ref{am-spectra}). This excess emission, which appears as a lower-velocity wing to the brighter \til50$-$55 \kms~velocity component, is detected in nearly all of our observed lines (\eg, \am, \cyano).  Since this excess is detected in multiple molecules and transitions, we {interpret it} to be an intermediate velocity component. Including a third component in our fitting process greatly reduced the residuals to produce the solid green residual in Figure \ref{am-spectra}. The final fit parameters used to produce the three Gaussian components in Figure \ref{am-spectra} are listed in Table \ref{55table}.

\begin{figure*}
\includegraphics[scale=0.5]{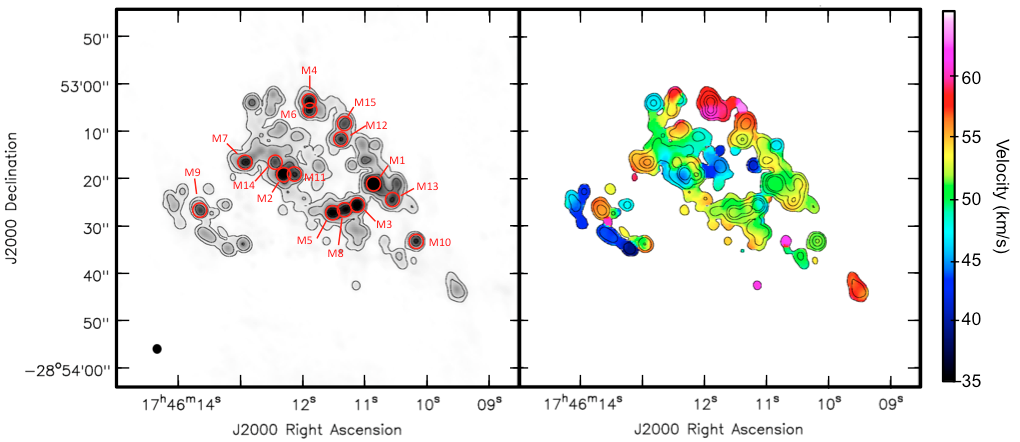}
\caption{Distribution of the 36.2 GHz \meth~(\mtrans) masers in \sticks~showing the maximum intensity emission ({\it left}), from Figure \ref{morphfig}, and central velocities ({\it right}) for emission above the 12$\sigma$ level.  The overlaid contours {show} 12, 30, 100, and 200 $\times$ 50 m\jyb~(rms noise in brightest channel). The 15 brightest masers from Table \ref{MaserTable} are marked on the \leftit~panel. }
\label{all maser}
\end{figure*}

The lowest velocity component, which has a central velocity of 10.6 \kms, is the faintest of the three components. This velocity component is detected in both the \cyano~transitions and in the \am~lower J-transitions (J$<$7). The highest velocity component, fit with a central velocity of 51.5 \kms, is the brightest of the three components and is detected in all of our observed molecular lines. This velocity component appears to dominate the moment 1 map, shown in Figure \ref{spectrum}. The intermediate velocity component, which is best fit with a central velocity of 37.6 \kms~in the \am~(5,5) line, is shown to be slightly spatially offset from the high velocity component in Figure \ref{spectrum}. We note that, while present, the central velocity of the intermediate velocity component did vary between the different molecular transitions, ranging from \til30$-$45 \kms. Therefore, the error estimates on the central velocity of the intermediate component are much larger than those shown in the Low and High velocity components to reflect this uncertainty.

We can further analyze the morphology of the molecular emission by isolating the channels associated with each component. Figure \ref{velocity-panels} shows the distribution of the \am~(3,3) emission in each velocity component, labeled by their respective central velocities from Table \ref{55table}. We are using the \am~(3,3) line for this analysis due to the {faintness} of the {low-velocity} component in the \am~(5,5) {transition}. When integrating over the field we were sensitive enough to detect the 10.6 \kms~component, but for a spatial mapping the \am~(5,5) line is not bright enough to perform a pixel by pixel analysis of {that} component. We are aware {that} the hyperfine {satellite lines of the} \am~(3,3) emission will be more prominent {than} in the (5,5) transition and will acknowledge where those lines may contribute in the following discussion.

In general, the observed gas morphology is unique for each velocity component. The 51.5 \kms~velocity gas is concentrated toward the center of the field and closely follows the bright \am~emission in Figure \ref{morphfig}, with the exception of the filamentary extension (\eg, see the wedge-shaped distribution in the red contour, Figure \ref{velocity-panels}). The 10.6 \kms~component is distributed throughout the field-of-view and contains several elongated structures (\eg, black box in Figure \ref{velocity-panels}). 
Further, this component does not appear to have similar morphology with the 51.5 \kms~component, suggesting this gas could be independent from the 51.5 \kms~emission. 
The morphology of the 37.6 \kms~component has similar attributes to both the 10.6 and 51.5 \kms~components. Unlike the 51.5 \kms~component, the 37.6 \kms~component $is$ associated with the filamentary extension. Further, the filamentary extension closely follows the elongated structure in the 10.6 \kms~component (Figure \ref{velocity-panels}). The 37.6 \kms~component also contains concentrated emission toward the north which spatially overlaps with emission in the 51.5 \kms~component. 
{Because of this spatial overlap,} some of the 37.6 \kms~emission could be from the hyperfine lines in the 51.5 \kms~component.


\subsection{36.2 GHz \meth~Masers in \sticks}
\label{masertext}

Our Ka-band observations included the 36.2 GHz CH$_3$OH (\mtrans) maser transition. This class I maser is known to trace shocks, as it is collisionally excited \citep{Morimoto85, Menten91a, Sj10}. The 36.2 GHz CH$_3$OH (\mtrans) maser transition has previously been detected towards this region \citep{YZ13, cotton16}. Our data {suggest} there are at least one hundred compact \meth~sources in \sticks~(Figure \ref{morphfig}, second row, last panel). The compact \meth~sources in \sticks~are located within a square arcminute region, and closely follow the bulk of the \am~and \cyano~emission at velocities from 40$-$60 \kms~(Figure \ref{velocity-panels}, bottom panel).  

{Figure \ref{all maser} shows the spatial distribution (left) and the velocity distribution (right) of the bright, above 0.6 \jybe~(12$\sigma$), 36.2 GHz \meth~emission. The \meth~emission is not uniformly distributed throughout {\sticks}. Most of the \meth~emission appears to be distributed throughout the wedge-like structure (discussed in Section \ref{g10morph}). We do not detect any compact emission, above 12$\sigma$, from the filamentary extension (e.g., see Figures \ref{morphfig} and \ref{all maser}).}

The velocity of the \meth~emission in M0.10$-$0.08 ranges from \til35 to 65 \kms. This corresponds to the velocity range of the bright \am~emission (Figure \ref{spectrum}). 
This velocity range indicates that most of the \meth~maser emission is associated with the 37.6 and 51.5 \kms~velocity components.

{In order to characterize the nature of the point-like emission and to evaluate whether these detections represent maser emission, we used the source detection algorithm, \clumpfind~\citep{Williams94}, to distinguish the emission both spectrally and spatially.~\clumpfind~identifies local maxima and uses saddle-points in position and velocity space around the local maxima to determine the boundaries of the sources. \clumpfind~then produces a list {of} clumps with uniform criteria, which was used to construct a catalog \citep[for more details on maser identification using the \clumpfind~algorithm, see the description of this technique in][{Section 5.1}]{mills15}. 
Sixty-four of the compact \meth~sources have brightness temperatures over 400 K (\ie, `CH$_3$OH masers').
The properties of the 64 detected `CH$_3$OH masers' 
identified with \clumpfind~are listed in Table \ref{MaserTable}. The spectral profiles of these~masers are shown in Figure \ref{spectra1}. The 15 brightest masers in \sticks~are labeled in Figure \ref{all maser} (left). }

With \clumpfind~we also detect 31 compact CH$_3$OH sources that have a brightness temperature between 100$-$400 K, {which we regard as `maser candidates.' 
These sources are considered to be candidate masers based on their brightness temperature, which are similar to observed gas temperatures in CMZ clouds \cite[50 - 400 K,][]{mills13, Krieger17} Therefore, we assume that any emission above this upper 400 K limit is likely non-thermal (i.e., maser emission, sources in Table \ref{MaserTable}) and any emission that is below 100 K is most likely thermal. Therefore, we classify \meth\ point-sources that have brightness temperatures between 100-400 K as `maser candidates'.}
The properties of all 31 maser candidates are listed in Table \ref{CanTable}, with their spectra shown in Figure \ref{spectra2}.  These maser candidates are also located within the same square arcminute region as the detected \meth~masers and have a similar velocity range (41$-$63 \kms).


\section{Discussion}
\label{dis}

In the following section we present a discussion and interpretation of our kinematics results on the \sticks~cloud (Section \ref{results}). Here, we attempt to explain the complicated and multiple-component velocity structure detected in the vicinity of \sticks~(Sections \ref{stream} \& \ref{g10-g11}). 

\subsection{Locations and Origins of \sticks~and \gecloud}
\label{stream}

The bulk emission in this region of the Galactic center has a velocity around 51.5 \kms~(Section \ref{3comp}). The morphology and gas kinematics of this velocity component {are consistent with those found in previous studies of \sticks~\citep[e.g.,][]{tsuboi11}}. 
\sticks~appears to be part of a larger structure of molecular gas that has a velocity of around 50 \kms~\citep{Fukui77,tsuboi11}. \cite{tsuboi11} detected H$^{13}$CO$^+$ emission around 50 \kms~{extending} from $+$0.15\degree~to $-$0.05\degree~($d$\til27 pc; see their Figure 10). Within this extended diffuse structure {they detect} three concentrated regions of H$^{13}$CO$^+$ emission that coincide with the \sticks, \stone, and {50 \kms~cloud (\fifty)} molecular clouds (see Figure \ref{introfig} for locations of these clouds). The presence of all three clouds within this larger diffuse structure could be evidence that all three clouds are co-located {within a single lower-density envelope} that has a velocity of around 50 \kms. 

This large diffuse gas structure, {observed in H$^{13}$CO$^+$ by \cite{tsuboi11},} may be Orbital Stream 1 in the \cite{Kru15} orbital model. The {50 \kms~cloud} is argued to be associated with Orbital Stream 1 \citep{Kru15}. Therefore, if \sticks~and \fifty~are associated within the same gas stream, and {50 \kms~cloud} is located on Orbital Stream 1, then by extension {we can infer that} \sticks~{is} also located on Orbital Stream 1. We note, however, that in the \cite{Kru15} orbital stream model, gas at the closest angular location to the \sticks~cloud ($l=$0\fdg09, $b=-$0\fdg07; as the position of \sticks~is slightly offset from Orbital Stream 1 by \til1\arcmin) is {predicted} to have a line-of-sight velocity of {60 to 65} \kms. Although this suggested line-of-sight velocity is slightly higher then the central velocity of \sticks~that we measured in Figure \ref{am-spectra}, we do detect some gas at velocities of around 60 to 65 \kms~(see Figures \ref{spectrum} \& \ref{all maser}, right).

The \mshell~is also hypothesized to be located on Orbital Stream 1 \citep[][{see our Figure \ref{introfig} for spatial location of shell relative to other GC clouds}]{my17}. In \cite{my17} we reported a systemic velocity of \til53 \kms~for the \scloud~expanding shell and advocate that the shell is also located on Orbital Stream 1, based on position-velocity analysis. Indeed, the adjacent locations of \sticks~and the \mshell~(see Figure \ref{introfig}) and their similar velocities are consistent with both clouds being
on the same orbital stream. Additionally, based on the orbital direction of stream 1, the \sticks~cloud would be located `upstream' from the \mshell. Based on the orbital solution in \cite{Kru15}, \sticks~{would} orbit into the current location of the \mshell~in \til0.05 Myrs. 

The 10.6 \kms~component (Figure \ref{am-spectra}; Section \ref{g10kin}) covers a velocity range of \til0$-$20 \kms~based on analysis of the \am~(5,5) emission and additional analysis of the \cyano~lines. This velocity range is similar to observed velocities of the adjacent \gecloud~molecular cloud \citep[\til10$-$30 \kms;][see their Figure 2]{Jones12,clavel13}, {and nearby gas velocities associated with with \cite{Kru15} orbital stream 3 (\til0--5 \kms)}. However, there are discrepancies in the literature concerning the velocity of the \gecloud~cloud. \cite{tsuboi97}, \cite{Handa06}, {and \cite{tsuboi11}} report a slightly higher velocity range of 15$-$45 \kms. These velocity values of the \gecloud~cloud in \cite{tsuboi97}, \cite{Handa06}, {and \cite{tsuboi11}} are closer to those of the intermediate velocity component in our observations (37.6 \kms; see Section \ref{3comp}). In our Figure \ref{am-spectra} this velocity component appears as a lower velocity `wing' of the main 51.5 \kms~component, rather than a distinct cloud. Further, the morphology of the 37.6 \kms~component in Figure \ref{velocity-panels} appears to overlap with the 51.5 \kms~component with the exception of the filamentary extension.  Therefore, based on the previous work of \cite{Jones12} and \cite{clavel13} and our analysis above, we interpret the 10.6 \kms~component as extended emission associated with \gecloud.


\subsubsection{Similar \xray~fluorescence detected in both \sticks~and \gecloud}
\label{xray-section}

Observed \xray~fluorescence can be beneficial in determining radial distances, which, when combined with their projected separation from the Galactic center, can be used to infer inter-cloud distances \citep[\eg,][]{clavel13, terrier18}. In our Galactic center, fluorescent iron emission at 6.4 keV is created in molecular clouds by K-shell photo-ionization and Compton scattering of neutral iron atoms from a previous, gigantic {\xray} flare, presumably from \sgra.
By observing the time delay of the detected \xray~reflection {across multiple molecular clouds}, we can constrain the locations of clouds from the geometrical path-length between the clouds and \sgra~and determine their location along our line-of-sight \citep[\eg,][]{CS02}. Further, the time delay between the detected reflections provides a measurement of the total path traveled by the photons, assuming they were emitted simultaneously.  This path length then gives an indication of the relative locations of the clouds. 

Molecular clouds that show similar illumination at a similar timeframe are located along the same {3-dimensional} `parabola', assuming the illumination feature is produced by the same, single flaring event \citep[\eg,][]{SC98}. Along this {3-dimensional} parabola, the path length of the propagating light signal (from \sgra~to the cloud and then to the Earth) is the same at each location and therefore the time delay of the propagating signal is the same as well. 
\cite{clavel13} detected a similar \xray~fluorescence signature in both \sticks~and \gecloud~(sources Br2 and G0.11$-$011 in their study).\footnote{The data presented in \cite{clavel13} used Chandra observations from 1999 to 2011 (see their Table 1 for observational information). These \xray~observations had a resolution of 4\arcsec, and are therefore fairly comparable to the observations presented in this paper, Table \ref{Images}.}
This detection of similar \xray~fluorescence illumination in \sticks~and \gecloud~indicates the two clouds are located along the same {3-dimensional} parabola, assuming the fluorescence in both clouds is from the same event. Further, because the two clouds are aligned along the same line of sight, and have a similar \xray~fluorescence {light curve}, \cite{clavel13} argue the two clouds must be at the same physical position, even with their differences in velocity. If the two clouds are almost at the same physical location, then we would expect to see evidence of this interaction.


\begin{figure}
\includegraphics[scale=0.32]{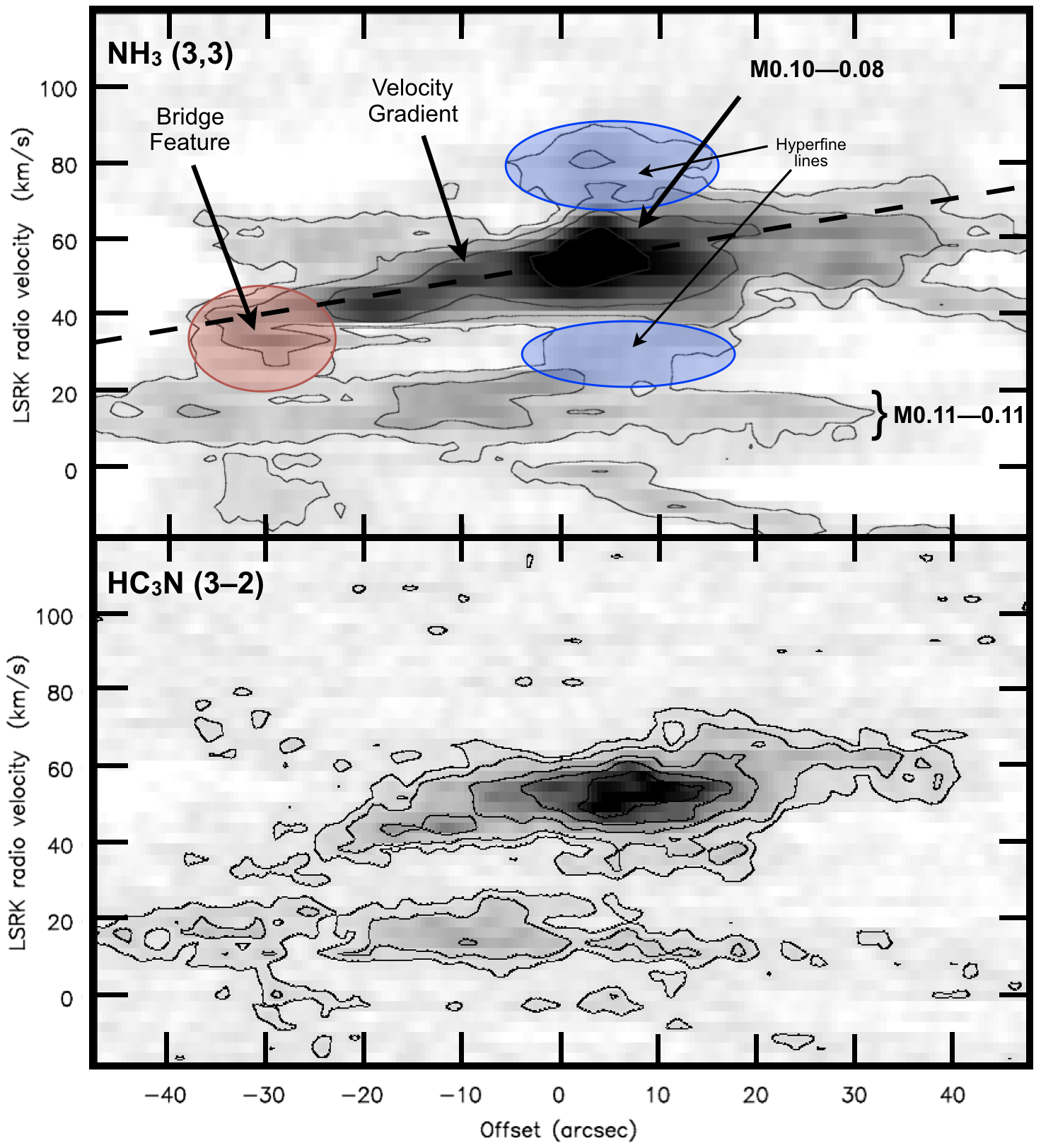} 
\caption{{Position-velocity distribution across the spatial slice shown in Figure \ref{velocity-panels}, for \am~(3,3) (\textit{top}) and \cyano~(3$-$2) (\textit{bottom}).  Annotations in the top panel show the gas associated with \gecloud~and \sticks~(see Section \ref{stream}). The black dashed line shows the magnitude and orientation of the \til10 \kmsp~velocity gradient described in Section \ref{g10kin}. The blue regions in the top panel show the general locations of the hyperfine satellite lines ($\pm$20$-$30 \kms~from the main component) of \sticks. The red shaded region highlights the `bridge'-like feature discussed in Section \ref{g10-g11}.  }}
\label{g10pv}
\end{figure}


\subsection{Proposed Physical Interaction between \sticks~and \gecloud}
\label{g10-g11}

Previous studies have hinted at a possible connection between \sticks~and \gecloud~\citep{Handa06, clavel13}. However, because of the large velocity difference between the two clouds {along this line-of-sight} ($\Delta$v\til 30 \kms), other investigators have suggested {these components} are physically separated \citep{ponti10,Kru15}. The high-resolution data presented in this paper can provide insight {into} this discrepancy in the literature. In this section we perform a detailed \pv~analysis on this region to investigate a possible connection between \sticks~and \gecloud.

Figure \ref{g10pv} shows the \pv~distribution of \am~(3,3) {(top) and \cyano~(3$-$2) (bottom)} across the filamentary extension (black box in Figure \ref{velocity-panels}). This slice was selected to maximize the relatively faint signal of the \gecloud~cloud (top panel in Figure \ref{velocity-panels}) and illustrate a possible connection to \sticks. 
The slice contains emission in all three velocity components (Table \ref{55table} and Figure \ref{velocity-panels}). Emission associated with the \sticks~cloud is clearly the brightest component in this region (50$-$60 \kms), with possible hyperfine lines above and below the main emission region (blue shaded region in Figure \ref{g10pv}, top). These hyperfine lines have a fixed known separation from the main component of $\pm$21.1 \kms~and $\pm$29.1 \kms~for the \am~(3,3) transition \citep[\eg,][]{Krieger17}. The \am~emission at \til80 \kms~is not observed in \cyano~{(3$-$2) (Figure \ref{g10pv}, bottom)}, suggesting it is hyperfine line emission. 
Across this slice there is clear, extended emission in \gecloud~(10.6 \kms~{component}; Figure \ref{g10pv}, top). The emission in \gecloud~is relatively faint compared to \sticks~and spans a velocity range from 5 to 25 \kms.

 The 37.6 \kms~component appears as a lower velocity wing to the 51.5 \kms~component in the integrated spectrum (Figure \ref{am-spectra}). When isolating velocity channels associated with each component we see that some of the gas in the 37.6 \kms~component is spatially offset from the 51.5 \kms~component (Figure \ref{velocity-panels}). 
 We also observe this offset {in} \pv~space, where some of the gas in the 37.6 \kms~component appears to be  spatially offset from the bulk of the 51.5 \kms~component (Figure \ref{g10pv}, top). Additionally, the 37.6 \kms~component is mainly associated with emission along the velocity gradient (see Section \ref{g10kin}) and appears to be a distinct feature in \pv~space. Further, at the southern edge of the velocity gradient there is a bridge feature with emission between velocities 20$-$40 \kms~(red shaded region in Figure \ref{g10pv}, top). Including both the bridge feature and the velocity gradient results in continuous emission between 20$-$50 \kms, thereby showing that \sticks~and \gecloud~have an apparent connection in \pv~space. Analysis of the \cyano~(3$-$2) line shows a similar velocity gradient and bridge-like features across the slice (Figure \ref{g10pv}, bottom). However, the \cyano~emission is \til5$-$10$\times$ fainter than the \am~(3,3) line, so these features appear loosely connected and barely above the noise level. 
 
 Recent studies simulating Galactic cloud-cloud interactions predict broad `bridge'-like feature in \pv~space, where the two clouds are physically connected \citep[\eg,][]{Takahira14, Haworth15a, Torii17}.\footnote{This labeling of the `bridge'-like feature in \pv~space, defined in \cite{Haworth15a}, should not be confused with the \xray~definition of the Bridge, labeled as Br1 and Br2 \citep{ponti10, clavel13}, which spatially connects \fifty~to \sticks.} In these studies, there is intermediate-velocity gas between the two main cloud components, which produces the `bridge' in position-velocity space. {Such} bridge features have since been detected in numerous molecular clouds throughout the galaxy \citep[\eg,][]{Fukui16, Torii17}. 
 
 Large-scale observations (\til45\arcsec~resolution) of the intermediate velocity component (15$-$45 \kms) from 
  \cite{tsuboi97} and \cite{Handa06} 
 show the gas is extended and dense. \cite{tsuboi97} extracted a position-velocity slice at $b$=$-$6\arcmin~(4\arcmin$\leq$$l$$\leq$14\arcmin) from their CS data cubes and observed two vertical features in velocity space that spanned 15 \kms~to 40 \kms, and were separated by \til2\arcmin~(see their Figure 2). \cite{Handa06} also saw similar vertical features in velocity space in their H$^{13}$CO$^+$ and SiO data cubes. 
 \cite{tsuboi97} attribute the vertical velocity features to be an expanding shell centered on a lower emission region near the centroid of the cloud, where the bright vertical features are the limb brightened edges of the shell. However, these vertical features could alternatively be signatures of the bridge feature, discussed above, on larger scales. 
 At the low spatial resolution of the \cite{Handa06} and \cite{tsuboi97} observations (\til45\arcsec), the detailed substructure we observed at -45\arcsec\ to 0\arcsec~in Figure \ref{g10pv} (top) 
 would blend into a single pixel. Therefore the high-resolution gradient and bridge features shown in Figure \ref{g10pv} would appear as broad, continuous emission at 45\arcsec~resolution.

Thus, based on the close physical proximity of M0.10$-$0.08 and \gecloud~from X-ray fluorescence data, along with continuous emission connecting them in position-velocity space via a `bridge' feature, we argue the two clouds {are} physical interacting. 
{Furthermore, this would imply that \gecloud~is located on the same stream as \sticks\ and not on a separate stream as indicated by the \cite{Kru15} orbital model.}

\subsection{Gas Kinematics in CMZ Clouds}

{Disentangling the molecular gas kinematics in CMZ clouds can be complex. As we have shown in this paper, the multiple velocity components toward the \sticks~cloud can make isolating the individual components challenging. 
For example, extensive efforts were conducted to fit the three components, with similar Gaussian fit parameters (V$_c$ \& $\sigma$), across multiple \am~transitions. However, we were unable to get converging values that satisfied the multiple transitions. The lower J transitions had brighter hyperfine structure for each component resulting in over nine blended profiles within the \til0$-$70 \kms~velocity range which could be fit with numerous solutions. At the higher J transitions, the 10.6 \kms~component was not bright enough to fit the spectrum.} 

{The 37.6 \kms~component was also especially challenging to fit. Since this component appears as a low-velocity wing to the 51.5 \kms~component, there were numerous solutions to the profile that varied depending on the initial guesses and range limits in the $pyspeckit$ program, with central velocity values that ranged from \til30$-$45 \kms. However, the presence of an intermediate velocity component between \til30$-$45 \kms~was clear in all of our \am~and \cyano~transitions (illustrated by the dashed green residuals in Figure \ref{am-spectra}). The fit solution we present in this paper (see Table \ref{55table}) was the best fit parameters that accurately reflected the uncertainty in the 37.6 \kms~component. However, we note that determining a simple kinematic solution to complex kinematics in the CMZ can be problematic and requires multiple methods to disentangle the velocity components (i.e., spectral fitting, position-velocity analysis, moment images of components, etc). }

{Furthermore, the complex kinematics in the CMZ can make understanding the gas flows and unusual orbits more challenging. We have attempted to disentangle the kinematics towards this complicated region using high spatial and spectral resolution observations. While we were able to identify the three velocity components towards this region using a variety of methods, providing a simple solution that satisfies the kinematics observed in this dataset is more difficult. Future observations of complex kinematic regions should use a variety of methods to isolate the velocity components. If possible, future observations should also use absorption observations toward radio continuum regions to constrain the line-of-sight arrangement, similar to the approach we used in \cite{my17}.   }
{Despite the complexity of disentangling the kinematics of multiple components, the analysis is necessary constrain models of the large-scale gas structures. Models for the 3-dimensional orientations of these gas structures can be influenced by assumptions made in complex kinematic regions. Therefore, applying the solutions in complex kinematic regions in the models may help resolve some of the contingencies in future orbital solutions. }


\section{Summary}
\label{conclusion}

We present high-resolution (\til3\arcsec) VLA radio observations of the {compact (3 pc)} M0.10$-$0.08 \mc, {finding that it is composed of} multiple compact molecular clumps (5+ clumps; D$_{clumps}$ $\leq$ 0.4 pc; Section \ref{mol-res}). We detect 15 molecular transitions in \sticks~(Table \ref{Images}); including eight transitions of \am, two \cyano~transitions, OCS, \methcy, \cyanobut, and abundant 36.2 GHz CH$_3$OH~masers (see Section \ref{masertext} and Appendix \ref{app} for details on the detected masers).

The main focus of this paper is on the molecular gas kinematics toward \sticks. We present the following results from this study: 

\textbf{1) \underline{Three velocity components detected toward} \underline{\sticks}:} {The averaged} \am~(5,5) spectrum {reveals} three velocity components {centered at} 10.6, 37.6, and 51.5 \kms~(see Section \ref{3comp}, Figures \ref{am-spectra} \& \ref{velocity-panels}, and Table \ref{55table}). 
{Initially, the \am~(5,5) spectrum was fit was two Gaussian components at \til10$-$15 \kms~and \til50$-$55 \kms. However, the residuals of this fit showed excess emission around 20$-$40 \kms, which we interpreted to be a third velocity component (see green dashed line in Figure \ref{am-spectra})}
In our high-resolution data the {51.5 \kms~component is the brightest emission in this region.}
The 10.6 \kms~component is relatively faint compared to the other two components in the field. We have also analyzed the gas morphology in each component by isolating channels associated with each component (Figure \ref{velocity-panels}). The morphology in all three components is unique.

\textbf{2) \underline{Relationship between \sticks~and Orb-} \underline{ital Stream 1}:} 
\sticks~is part of a larger structure of gas that contains the \fifty~and \stone~molecular clouds and has a velocity of around 50 \kms~\citep{tsuboi11}. 
The central velocity of M0.10$-$0.08 (51.5 \kms; Section \ref{stream}) indicates that M0.10$-$ 0.08 is likely located on Orbital Stream 1 in the \cite{Kru15} model. 

\textbf{3)~\underline{Resolving the Kinematics of \gecloud}:} Discrepant reports regarding the central velocity of M0.11$-$ 0.11 range from 10 to 45 \kms. In our high-resolution data, we detect two components in this velocity range: 10.6 and 37.6 \kms. We argue that gas in the 10.6 \kms~component is associated with \gecloud~as the morphology is distinct from that of the \sticks~cloud (Figure \ref{velocity-panels}). Additionally, \pv~analysis towards this region of the CMZ shows extended emission ($>$70\arcsec; $>$2.7 pc) from 0 to 20 \kms~(Figure \ref{g10pv}), that we suggest is associated with \gecloud.

\textbf{4)~\underline{Physical Interaction Between \sticks} \underline{and \gecloud}:} Past X-ray fluorescence observations by \cite{clavel13} show similar time delay signatures from both \sticks~and \gecloud~and argue the two clouds are in the same physical position of the Galactic center. 
The intermediate morphology of the 37.6 \kms~velocity component could be indicative of a physical interaction between \sticks~and \gecloud. Indeed, all three velocity components appear to be connected in \pv~space (Figure \ref{g10pv}). The intermediate velocity component, which has similar features to both \sticks~and \gecloud, could be gas from where these two are physically connected.


\section{Acknowledgements}

This material is based upon work supported by grants from the National Science Foundation (NSF; no. AST-0907934, AST-15243000, AST-1614782, AST-2008101, CAREER-2142300). Support for this work was also provided by the NSF through the Grote Reber Fellowship Program administered by Associated Universities, Inc./National Radio Astronomy Observatory. 

NOB would like to thank Dr. Farhad Yusef-Zadah {(Northwestern University)}, Dr. Robert Mutel {(University of Iowa)}, Dr. Steven Spangler {(University of Iowa)}, and Dr. Kennith Gayley {(University of Iowa)} for their helpful critique on this thesis analysis. The authors would also like to thank Dr. Allison Costa (University of Virginia), Dr. Monica Sanchez (University of New Mexico), and Dr. Diederik Kruijssen (Max-Planck Institute) for their helpful insight on this work. 

The authors would like to thank Dr. Elisabeth Mills {(University of Kansas)} for her helpful insight on the spectral line fitting and analysis of the results presented in this paper. The authors would also like to thank Dr. John Bally for his inspiration in creating the 3-color image shown in Figure \ref{introfig}. ﻿
{The authors would also like to thank the anonymous referee for their helpful insight on this manuscript.}

\software{CASA \citep{2011ascl.soft07013I}; \clumpfind~\citep{2011ascl.soft07014W}; $pyspeckit$ \citep{2011ascl.soft09001G,ginsburg22}}


%

\appendix
\counterwithin{figure}{section}
\counterwithin{table}{section}


\section{Catalogue of \meth~masers in \sticks}
\label{app}


To catalogue the properties of these masers we used the \clumpfind~algorithm \citep{Williams94}. We define all compact sources with peak brightness temperatures above 400 K as \meth~masers in this paper, following the classification used in \cite{mills15}. 
Table \ref{MaserTable} presents the properties of these 64 masers, with their spectra shown in Figure \ref{spectra1}. Table \ref{CanTable} lists the maser candidates with their spectra shown in Figure \ref{spectra2}. Results of all 95 detected compact \meth~sources (both masers and candidate masers) are discussed in Section \ref{masertext}.

\centering
\begin{longtable}{cccccccccc}
\caption{\\ \sc{~36.2 GHz \meth~Masers in \sticks}}
\\
\hline\hline
{\bf ID} & {\bf Maser	Name}&	{\bf $\alpha$}&	{\bf $\delta$}&	{\bf $v$}&	{\bf FWHM}&	{\bf I$_{peak}$}&	 {\bf Flux}&	{\bf T$_{b}$}&	{\bf Resolved?}\\
& & (J2000)	 & (J2000) & 	\kms & 	\kms	 & \jyb & 	Jy \kms	& K	 & \\
\hline 
\hline 
  M1&   M0.1039644-0.0802624   &17\h46\m10\fs85   &-28\degr53\arcmin21\farcs1   & 50.60   & 3.911   & 46.969   &265.944&    12438   &YES \\ 
  M2&   M0.1072762-0.0845517   &17\h46\m12\fs33   &-28\degr53\arcmin18\farcs9   & 43.25   & 7.833   & 29.621   &252.342&     7844   &YES \\  
  M3&   M0.1033757-0.0817884   &17\h46\m11\fs12   &-28\degr53\arcmin25\farcs7   & 46.40   & 4.332   & 24.917   &100.982&     6598   & NO \\  
  M4&   M0.1100239-0.0810107   &17\h46\m11\fs89   &-28\degr53\arcmin03\farcs8   & 59.00   & 3.289   & 20.721   & ~89.420&     5487   &YES \\ 
  M5&   M0.1038130-0.0832737   &17\h46\m11\fs53   &-28\degr53\arcmin27\farcs2   & 52.70   & 4.527   & 19.008   &131.418&     5033   &YES \\  
  M6&   M0.1096832-0.0812181   &17\h46\m11\fs89   &-28\degr53\arcmin05\farcs3   & 61.10   & 3.197   & 16.980   & ~70.044&     4496   &YES \\ 
  M7&   M0.1090134-0.0860631   &17\h46\m12\fs93   &-28\degr53\arcmin16\farcs4   & 55.85   & 4.772   & 16.369   &141.360&     4334   &YES \\  
  M8&   M0.1035684-0.0824885   &17\h46\m11\fs32   &-28\degr53\arcmin26\farcs4   & 51.65   & 4.394   & 15.666   &108.233&     4148   &YES \\ 
  M9&   M0.1078915-0.0897819   &17\h46\m13\fs64   &-28\degr53\arcmin26\farcs8   & 54.80   & 2.835   & 15.543   & ~62.866&     4116   & NO \\ 
 M10&   M0.0997717-0.0798958   &17\h46\m10\fs17   &-28\degr53\arcmin33\farcs3   & 50.60   & 2.093   & 13.577   & 41.198&     3595   & NO \\ 
 M11&   M0.1068280-0.0840072   &17\h46\m12\fs14   &-28\degr53\arcmin19\farcs3   & 51.65   & 6.166   & 11.870   &133.134&     3143   &YES \\
 M12&   M0.1072164-0.0806180   &17\h46\m11\fs40   &-28\degr53\arcmin11\farcs7   & 51.65   & 4.871   & 11.623   & 82.500&     3078   &YES \\ 
 M13&   M0.1026273-0.0797920   &17\h46\m10\fs55   &-28\degr53\arcmin24\farcs3   & 54.80   & 5.653   & 11.497   &121.184&     3044   &YES \\
 M14&   M0.1081318-0.0846147   &17\h46\m12\fs46   &-28\degr53\arcmin16\farcs4   & 55.85   & 5.166   & 10.121   & 82.795&     2680   &YES \\ 
 M15&   M0.1078793-0.0799809   &17\h46\m11\fs34   &-28\degr53\arcmin08\farcs5   & 55.85   & 4.417   & 10.026   & 70.744&     2655   &YES \\ 
 M16&   M0.1117873-0.0839075   &17\h46\m12\fs82   &-28\degr53\arcmin03\farcs8   & 46.40   & 3.953   &  9.862   & 48.569&     2611   & NO \\ 
 M17&   M0.1032901-0.0791549   &17\h46\m10\fs50   &-28\degr53\arcmin21\farcs1   & 52.70   & 5.703   &  9.169   & 86.949&     2428   &YES \\
 M18&   M0.1054681-0.0800475   &17\h46\m11\fs02   &-28\degr53\arcmin16\farcs0   & 52.70   & 6.421   &  7.465   & 55.618&     1977   &YES \\
 M19&   M0.1048914-0.0886890   &17\h46\m12\fs96   &-28\degr53\arcmin34\farcs0   & 55.85   & 3.140   &  6.689   & 23.222&     1771   & NO \\ 
 M20&   M0.0961419-0.0793034   &17\h46\m09\fs51   &-28\degr53\arcmin43\farcs3   & 57.95   & 9.415   &  6.398   & 75.184&     1694   &YES \\
 M21&   M0.1090096-0.0854816   &17\h46\m12\fs79   &-28\degr53\arcmin15\farcs3   & 50.60   & 7.226   &  5.984   & 61.843&     1584   &YES \\
 M22&   M0.1089767-0.0909895   &17\h46\m14\fs08   &-28\degr53\arcmin25\farcs7   & 44.30   & 4.955   &  5.751   & 51.748&     1522   &YES \\
 M23&   M0.1090577-0.0849852   &17\h46\m12\fs68   &-28\degr53\arcmin14\farcs2   & 49.55   & 6.802   &  5.465   & 58.627&     1447   &YES \\
 M24&   M0.1115057-0.0826777   &17\h46\m12\fs49   &-28\degr53\arcmin02\farcs4   & 60.05   & 2.935   &  5.460   & 22.068&     1445   & NO \\ 
 M25&   M0.1066804-0.0902855   &17\h46\m13\fs59   &-28\degr53\arcmin31\farcs5   & 44.30   & 4.414   &  5.459   & 40.981&     1445   &YES \\
 M26&   M0.1034462-0.0820959   &17\h46\m11\fs21   &-28\degr53\arcmin26\farcs1   & 59.00   & 5.071   &  5.279   & 38.491&     1398   &YES \\
 M27&   M0.1084206-0.0843221   &17\h46\m12\fs44   &-28\degr53\arcmin15\farcs0   & 48.50   & 6.924   &  5.277   & 77.437&     1397   &YES \\
 M28&   M0.1051877-0.0895595   &17\h46\m13\fs20   &-28\degr53\arcmin34\farcs7   & 39.05   & 2.862   &  5.217   & 18.355&     1381   &YES \\
 M29&   M0.1101091-0.0809589   &17\h46\m11\fs89   &-28\degr53\arcmin03\farcs5   & 54.80   & 3.356   &  5.175   & 31.380&     1370   &YES \\ 
 M30&   M0.1062866-0.0796660   &17\h46\m11\fs04   &-28\degr53\arcmin12\farcs8   & 52.70   & 5.321   &  4.661   & 36.543&     1234   &YES \\ 
 M31&   M0.0958678-0.0792367   &17\h46\m09\fs46   &-28\degr53\arcmin44\farcs0   & 53.75   & 3.893   &  4.447   & 20.439&     1177   & NO \\ 
 M32&   M0.1096798-0.0833221   &17\h46\m12\fs38   &-28\degr53\arcmin09\farcs2   & 59.00   & 3.828   &  4.372   & 26.852&     1157   & NO \\ 
 M33&   M0.1107909-0.0832296   &17\h46\m12\fs52   &-28\degr53\arcmin05\farcs6   & 54.80   & 4.924   &  4.322   & 29.242&     1144   & NO \\ 
 M34&   M0.1095427-0.0832887   &17\h46\m12\fs36   &-28\degr53\arcmin09\farcs6   & 52.70   & 4.907   &  4.176   & 46.350&     1105   &YES \\
 M35&   M0.1064915-0.0901670   &17\h46\m13\fs53   &-28\degr53\arcmin31\farcs8   & 41.15   & 6.140   &  4.170   & 35.602&     1104   &YES \\
 M36&   M0.1022351-0.0825996   &17\h46\m11\fs15   &-28\degr53\arcmin30\farcs8   & 52.70   & 5.473   &  3.974   & 49.319&     1052   &YES \\
 M37&   M0.1065278-0.0825552   &17\h46\m11\fs75   &-28\degr53\arcmin17\farcs5   & 44.30   & 6.437   &  3.715   & 33.159&      983   &YES \\ 
 M38&   M0.1019943-0.0823959   &17\h46\m11\fs07   &-28\degr53\arcmin31\farcs1   & 50.60   & 4.645   &  3.549   & 28.575&      939   &YES \\ 
 M39&   M0.1093989-0.0901487   &17\h46\m13\fs94   &-28\degr53\arcmin22\farcs9   & 42.20   & 4.945   &  3.496   & 25.923&      925   &YES \\ 
 M40&   M0.1111797-0.0825258   &17\h46\m12\fs41   &-28\degr53\arcmin03\farcs1   & 43.25   & 3.953   &  3.494   & 12.724&      925   & NO \\ 
 M41&   M0.1114205-0.0827295   &17\h46\m12\fs49   &-28\degr53\arcmin02\farcs8   & 55.85   & 4.550   &  3.452   & 32.194&      914   & NO \\ 
 M42&   M0.1058716-0.0789844   &17\h46\m10\fs82   &-28\degr53\arcmin12\farcs8   & 46.40   & 4.554   &  3.103   & 19.923&      821   &YES \\
 M43&   M0.1074210-0.0884336   &17\h46\m13\fs26   &-28\degr53\arcmin25\farcs7   & 47.45   & 1.695   &  3.022   &  7.059&      800   & NO \\ 
 M44&   M0.1055756-0.0807994   &17\h46\m11\fs21   &-28\degr53\arcmin17\farcs1   & 41.15   & 3.012   &  2.965   &  9.531&      785   &YES \\
 M45&   M0.1087352-0.0827294   &17\h46\m12\fs11   &-28\degr53\arcmin11\farcs0   & 48.50   & 3.934   &  2.695   & 26.288&      713   &YES \\ 
 M46&   M0.1074951-0.0893226   &17\h46\m13\fs48   &-28\degr53\arcmin27\farcs2   & 50.60   & 2.287   &  2.534   &  5.071&      671   & NO \\ 
 M47&   M0.1083131-0.0835702   &17\h46\m12\fs25   &-28\degr53\arcmin13\farcs9   & 48.50   & 8.930   &  2.459   & 32.152&      651   &YES \\ 
 M48&   M0.1044687-0.0841590   &17\h46\m11\fs84   &-28\degr53\arcmin26\farcs8   & 51.65   & 4.615   &  2.431   & 20.260&      643   &YES \\ 
 M49&   M0.1050284-0.0887224   &17\h46\m12\fs98   &-28\degr53\arcmin33\farcs6   & 46.40   & 7.745   &  2.388   & 27.822&      632   &YES \\ 
 M50&   M0.1079693-0.0912524   &17\h46\m14\fs00   &-28\degr53\arcmin29\farcs3   & 42.20   & 4.190   &  2.353   & 22.536&      623   &YES \\ 
 M51&   M0.1090804-0.0911599   &17\h46\m14\fs13   &-28\degr53\arcmin25\farcs7   & 40.10   & 3.113   &  2.304   & 11.128&      610   & NO \\ 
 M52&   M0.0995238-0.0812144   &17\h46\m10\fs44   &-28\degr53\arcmin36\farcs5   & 50.60   & 3.536   &  2.292   & 16.356&      606   &YES \\
 M53&   M0.1026091-0.0822551   &17\h46\m11\fs12   &-28\degr53\arcmin29\farcs0   & 45.35   & 4.354   &  2.184   & 13.540&      578   & NO \\ 
 M54&  M0.1087914-0.0887671   &17\h46\m13\fs53   &-28\degr53\arcmin22\farcs1   & 52.70   & 1.771   &  2.105   &  4.209&      557   & NO \\ 
 M55&   M0.1099096-0.0844667   &17\h46\m12\fs68   &-28\degr53\arcmin10\farcs7   & 48.50   & 5.256   &  2.009   & 26.867&      531   &YES \\ 
 M56&   M0.1071396-0.0898893   &17\h46\m13\fs56   &-28\degr53\arcmin29\farcs3   & 60.05   & 2.557   &  1.847   &  5.353&      489   & NO \\
 M57&   M0.1055655-0.0897965   &17\h46\m13\fs31   &-28\degr53\arcmin34\farcs0   & 42.20   & 6.025   &  1.838   & 18.726&      486   &YES \\
 M58&   M0.1105910-0.0840519   &17\h46\m12\fs68   &-28\degr53\arcmin07\farcs8   & 52.70   & 4.022   &  1.737   & 18.043&      460   &YES \\
 M59&   M0.1017160-0.0790623   &17\h46\m10\fs25   &-28\degr53\arcmin25\farcs7   & 51.65   & 2.259   &  1.690   &  9.070&      447   &YES \\ 
 M60&   M0.1061836-0.0848665   &17\h46\m12\fs25   &-28\degr53\arcmin22\farcs9   & 53.75   & 4.861   &  1.662   & 13.114&      440   & NO \\ 
 M61&   M0.1053243-0.0842220   &17\h46\m11\fs97   &-28\degr53\arcmin24\farcs3   & 54.80   & 3.553   &  1.640   & 11.119&      434   &YES \\
 M62&   M0.1007572-0.0815144   &17\h46\m10\fs69   &-28\degr53\arcmin33\farcs3   & 62.15   & 2.780   &  1.592   &  5.730&      421   & NO \\
 M63&   M0.1094015-0.0799883   &17\h46\m11\fs56   &-28\degr53\arcmin03\farcs8   & 56.90   & 7.673   &  1.539   & 17.278&      407   &YES \\
 M64&   M0.0068760-0.0835109   &17\h46\m12\fs03   &-28\degr53\arcmin18\farcs2   & 43.25   & 6.249   &  1.538   & 26.383&      407   &YES \\
  \hline
   \hline
 \label{MaserTable}
\end{longtable}
  

 \centering
\begin{longtable}{cccccccccc}
\caption{\\ \sc{36.2 GHz \meth~Maser Candidates in \sticks}} 
\\ 
\hline\hline
{\bf ID} & {\bf Maser	Name} &	{\bf $\alpha$}&	{\bf $\delta$}&	{\bf $v$}&	{\bf FWHM}&	{\bf I$_{peak}$}&	 {\bf Flux}&	{\bf T$_{b}$}&	{\bf Resolved?}\\
& & (J2000)	 & (J2000) & 	\kms & 	\kms	 & \jyb & 	Jy \kms	& K	 & \\ 
\hline 
\hline
 CM1 &   M0.1017240-0.0829108   &17\h46\m11\fs15   &-28\degr53\arcmin32\farcs9   & 55.85   & 6.634   &  1.498    & 20.043&      396   &YES \\ 
 CM2 &   M0.1073500-0.0827553   &17\h46\m11\fs92   &-28\degr53\arcmin15\farcs3   & 45.35   & 6.884   &  1.436   & 17.249&      380   &YES \\ 
 CM3 &   M0.1087496-0.0796846   &17\h46\m11\fs40   &-28\degr53\arcmin05\farcs3   & 63.20   & 6.767   &  1.418   & 17.226&      375   &YES \\
 CM4 &   M0.1057278-0.0831589   &17\h46\m11\fs78   &-28\degr53\arcmin21\farcs1   & 48.50   & 8.387   &  1.366   & 14.389&      361   &YES \\ 
 CM5 &   M0.1114133-0.0842520   &17\h46\m12\fs85   &-28\degr53\arcmin05\farcs6   & 50.60   & 3.705   &  1.365    &  8.593&      361   &YES \\ 
 CM6 &   M0.0994242-0.0843109   &17\h46\m11\fs15   &-28\degr53\arcmin42\farcs6   & 61.10   & 2.678   &  1.333   &  2.716&      352   & NO \\ 
 CM7 &   M0.1073396-0.0890670   &17\h46\m13\fs39   &-28\degr53\arcmin27\farcs2   & 44.30   & 1.908   &  1.323   &  2.607&      350   & NO \\ 
 CM8 &   M0.1061099-0.0893484   &17\h46\m13\fs28   &-28\degr53\arcmin31\farcs5   & 52.70   & 2.900   &  1.252   &  5.884&      331   &YES \\
 CM9 &   M0.1073796-0.0820367   &17\h46\m11\fs75   &-28\degr53\arcmin13\farcs9   & 41.15   & 6.005   &  1.210   & 10.429&      320   & NO \\
 CM10 &   M0.1069128-0.0812699   &17\h46\m11\fs51   &-28\degr53\arcmin13\farcs9   & 46.40   & 6.136   &  1.199   &  9.117&      317   &YES \\ 
 CM11 &   M0.1054022-0.0856925   &17\h46\m12\fs33   &-28\degr53\arcmin26\farcs8   & 53.75   & 1.864   &  1.165   &  3.439&      308   & NO \\ 
 CM12 &   M0.1037539-0.0847109   &17\h46\m11\fs86   &-28\degr53\arcmin30\farcs0   & 53.75   & 2.552   &  1.063   &  6.657&      281   &YES \\
 CM13 &   M0.1025205-0.0844109   &17\h46\m11\fs62   &-28\degr53\arcmin33\farcs3   & 53.75   & 1.525   &  1.020   &  1.581&      270   & NO \\ 
 CM14 &   M0.1086098-0.0871261   &17\h46\m13\fs12   &-28\degr53\arcmin19\farcs6   & 59.00   & 8.039   &  0.989   &  8.824&      261   &YES \\
 CM15 &   M0.1034571-0.0811550   &17\h46\m10\fs99   &-28\degr53\arcmin24\farcs3   & 56.90   & 5.577   &  0.987    &  9.074&      261   &YES \\  
 CM16 &   M0.0998050-0.0797587   &17\h46\m10\fs14   &-28\degr53\arcmin32\farcs9   & 56.90   & 3.739   &  0.945   &  5.708&      250   & NO \\ 
 CM17 &   M0.1070237-0.0799179   &17\h46\m11\fs21   &-28\degr53\arcmin11\farcs0   & 60.05   & 7.663   &  0.930   & 11.283&      246   &YES \\
 CM18 &   M0.1080322-0.0903968   &17\h46\m13\fs80   &-28\degr53\arcmin27\farcs5   & 41.15   & 4.784   &  0.889   &  7.060&      235   &YES \\
 CM19 &   M0.1053015-0.0807328   &17\h46\m11\fs15   &-28\degr53\arcmin17\farcs8   & 45.35   & 4.598   &  0.860   &  7.262&      227   &YES \\
 CM20 &   M0.0989385-0.0806366   &17\h46\m10\fs22   &-28\degr53\arcmin37\farcs2   & 53.75   & 5.362   &  0.855    &  6.528&      226   &YES \\  
 CM21 &   M0.1002794-0.0816885   &17\h46\m10\fs66   &-28\degr53\arcmin35\farcs1   & 49.55   & 2.898   &  0.824   &  6.111&      218   &YES \\  
 CM22 &   M0.1051058-0.0848221   &17\h46\m12\fs08   &-28\degr53\arcmin26\farcs1   & 52.70   & 3.773   &  0.819    &  5.464&      216   &YES \\
 CM23 &   M0.1093610-0.0816478   &17\h46\m11\fs94   &-28\degr53\arcmin07\farcs1   & 55.85   & 4.041   &  0.811    &  4.470&      214   &YES \\ 
 CM24 &   M0.1043426-0.0831849   &17\h46\m11\fs59   &-28\degr53\arcmin25\farcs4   & 45.35   & 8.092   &  0.730   &  5.279&      193   &YES \\ 
 CM25 &   M0.1015946-0.0840405   &17\h46\m11\fs40   &-28\degr53\arcmin35\farcs4   & 48.50   & 2.788   &  0.678   &  2.904&      179   &YES \\
 CM26 &   M0.1090351-0.0814959   &17\h46\m11\fs86   &-28\degr53\arcmin07\farcs8   & 52.70   & 4.308   &  0.668    &  2.702&      176   & NO \\
 CM27 &   M0.1083018-0.0818256   &17\h46\m11\fs84   &-28\degr53\arcmin10\farcs6   & 45.35   & 4.275   &  0.586   &  4.017&      155   & NO \\ 
 CM28 &   M0.1074053-0.0807365   &17\h46\m11\fs45   &-28\degr53\arcmin11\farcs4   & 47.45   & 4.038   &  0.572   &  3.719&      151   &YES \\
 CM29 &   M0.1016794-0.0813032   &17\h46\m10\fs77   &-28\degr53\arcmin30\farcs0   & 48.50  & $<$1.05   &  0.540   &  0.553&      143   & NO \\
 CM30 &   M0.1987015-0.0810144   &17\h46\m10\fs28   &-28\degr53\arcmin38\farcs7   & 43.25   & $<$1.05   &  0.504   &  0.365&      133   & NO \\
 CM31 &   M0.1044869-0.0816958   &17\h46\m11\fs26   &-28\degr53\arcmin22\farcs1   & 62.15   & 4.628   &  0.421   &  4.661&      111   &YES \\ 
 \hline
\hline
\label{CanTable}
\end{longtable}

\clearpage

\centering
\begin{figure*}
\centering 
\includegraphics[scale=0.83]{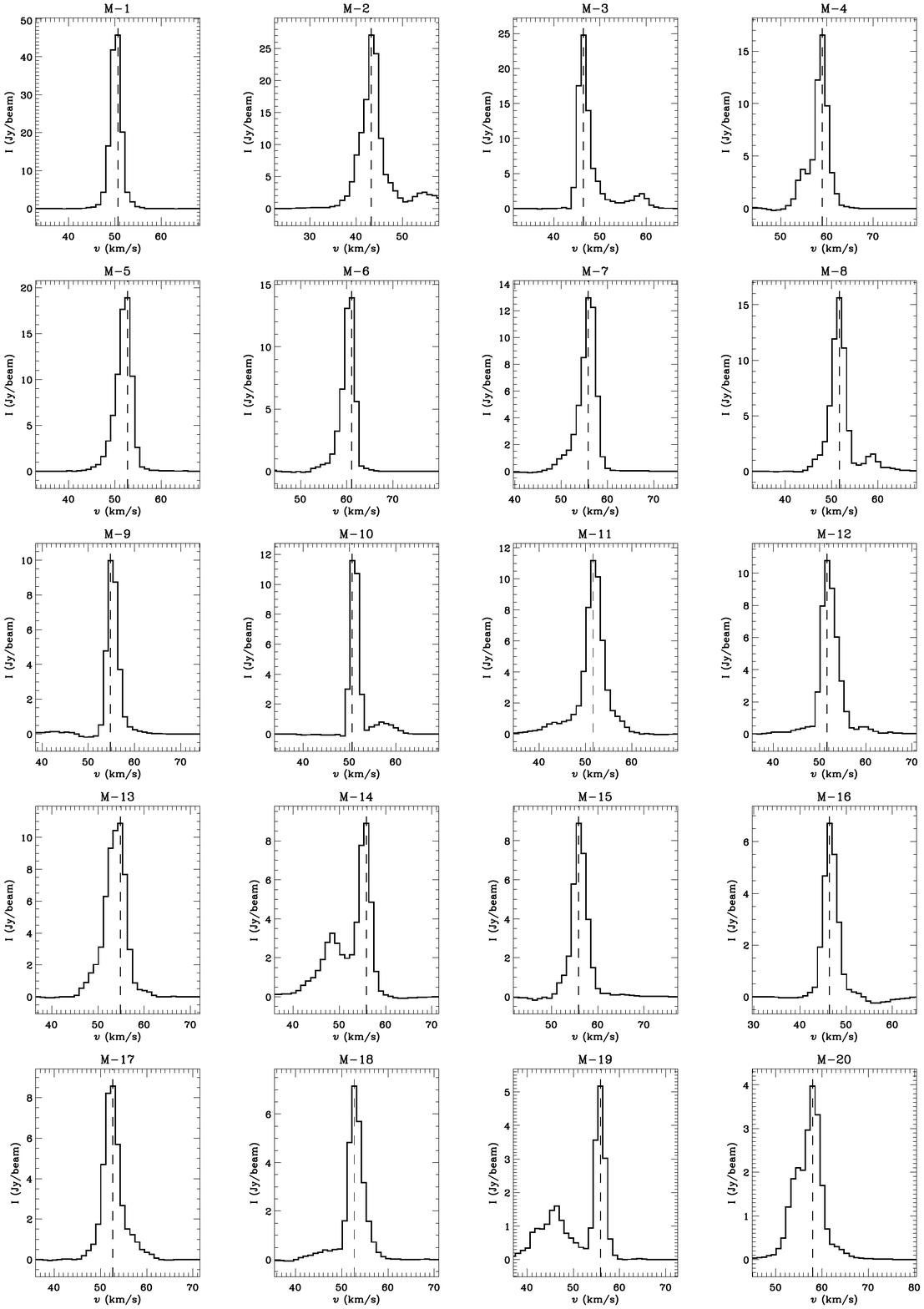} 
\caption{Spectra of detected 36 GHz \meth~masers in \sticks. }
\label{spectra1}
\end{figure*}

\renewcommand{\thefigure}{A.\arabic{figure}}
\addtocounter{figure}{-1}

\begin{figure*}
\includegraphics[scale=0.83]{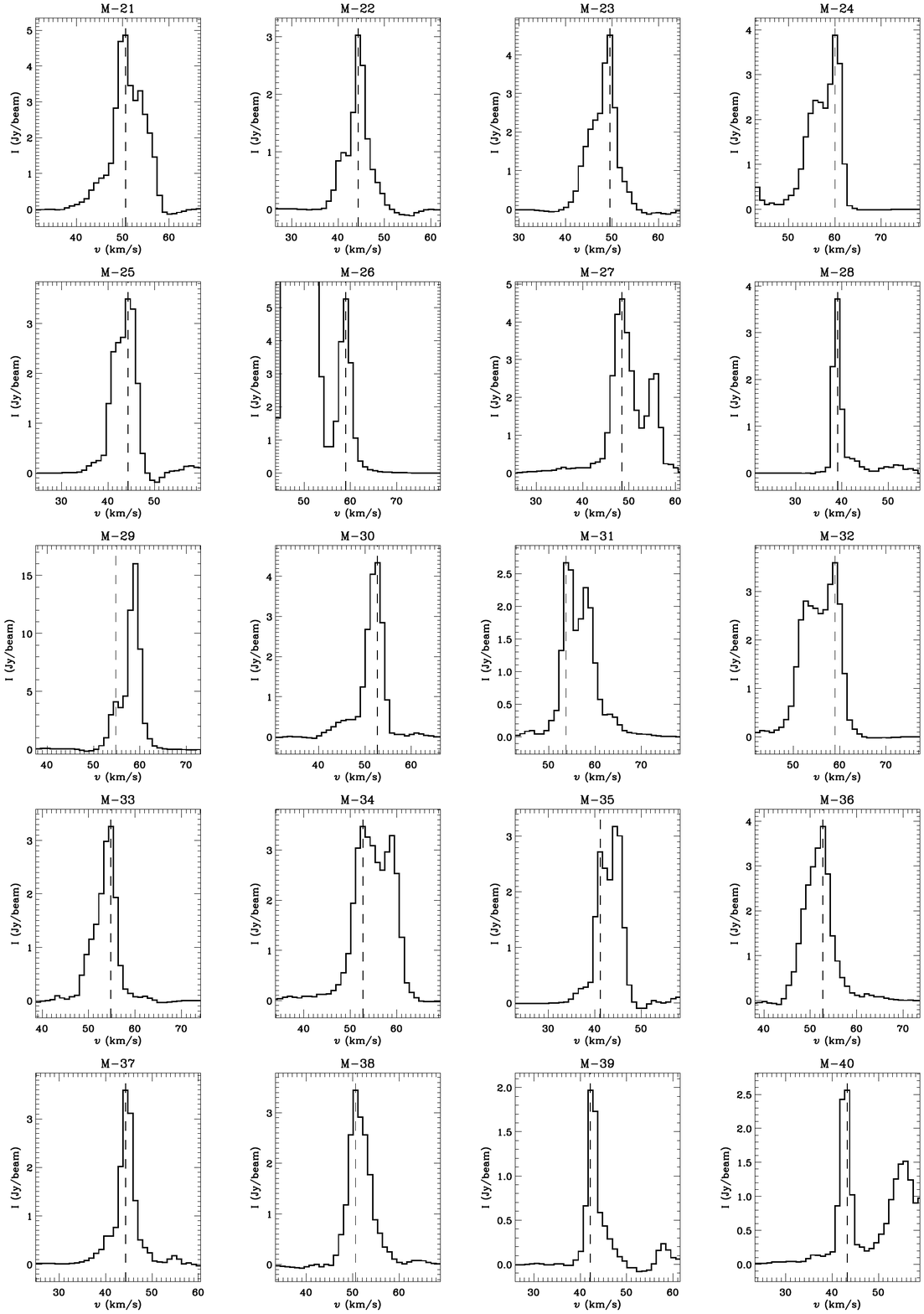} 
\caption{(Continued)}
\end{figure*}

\renewcommand{\thefigure}{A.\arabic{figure}}
\addtocounter{figure}{-1}

\begin{figure*}
\includegraphics[scale=0.83]{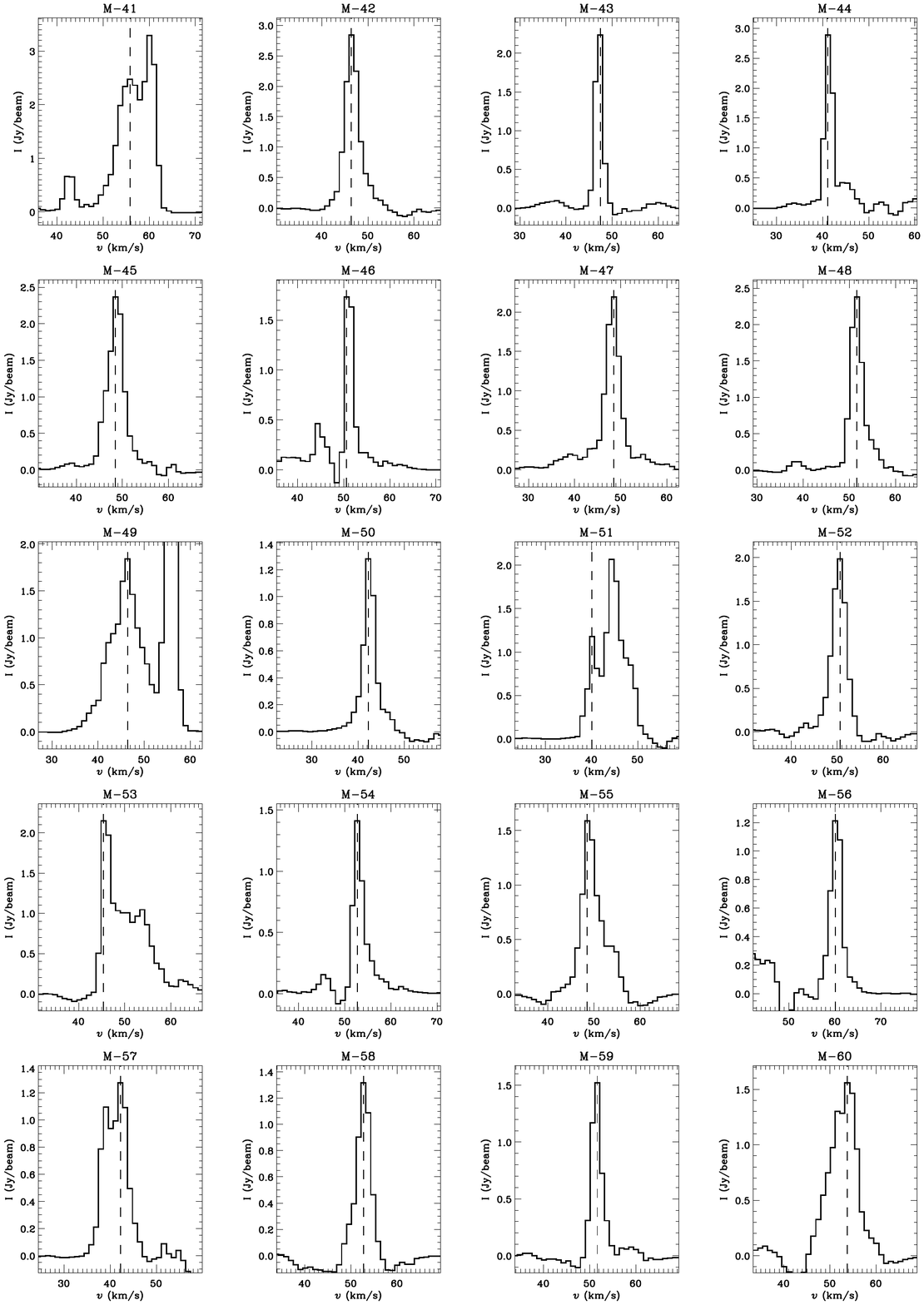} 
\caption{(Continued)}
\end{figure*}

\renewcommand{\thefigure}{A.\arabic{figure}}
\addtocounter{figure}{-1}

\clearpage

\begin{figure*}[t!]
\includegraphics[scale=0.83]{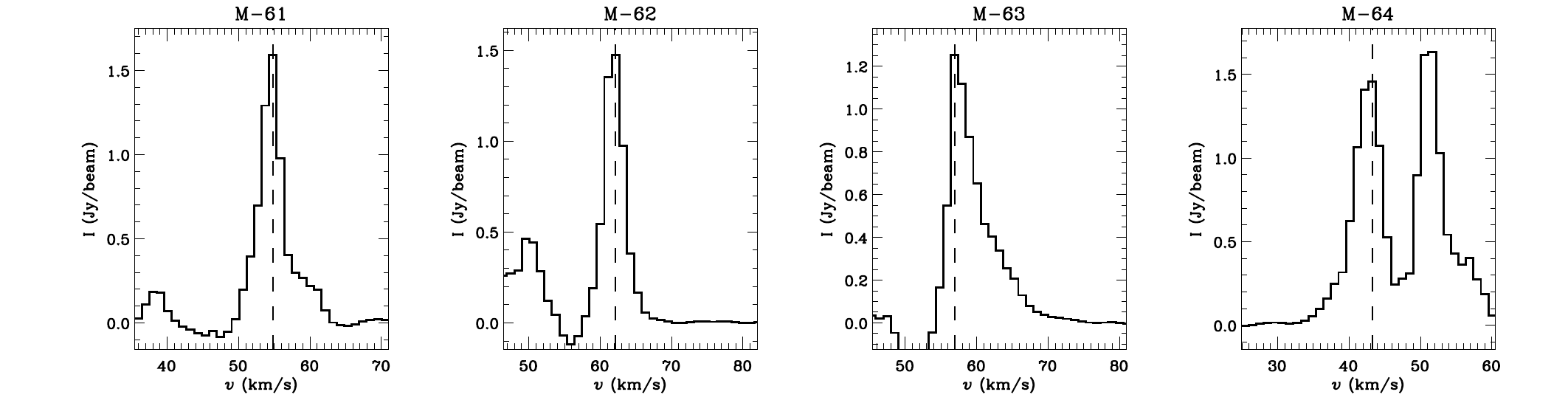} 
\caption{(Continued)}
\end{figure*}


\renewcommand{\thefigure}{A.\arabic{figure}}


\begin{figure*}
\includegraphics[scale=0.83]{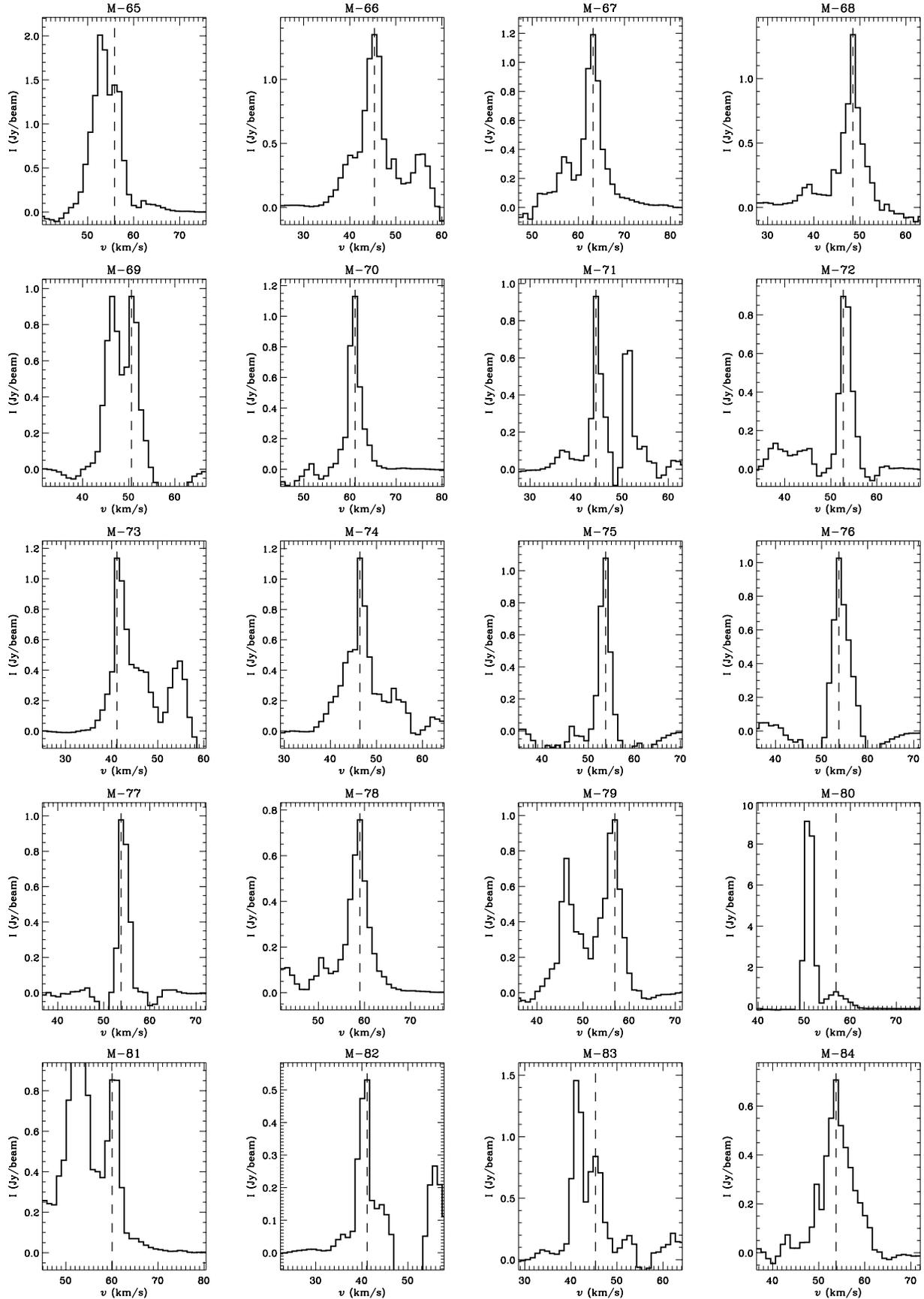} 
\caption{maser candidates CM1(M65) to CM16(M60)}
\label{spectra2}
\end{figure*}

\renewcommand{\thefigure}{A.\arabic{figure}}
\addtocounter{figure}{-1}

\begin{figure*}
\includegraphics[scale=0.83]{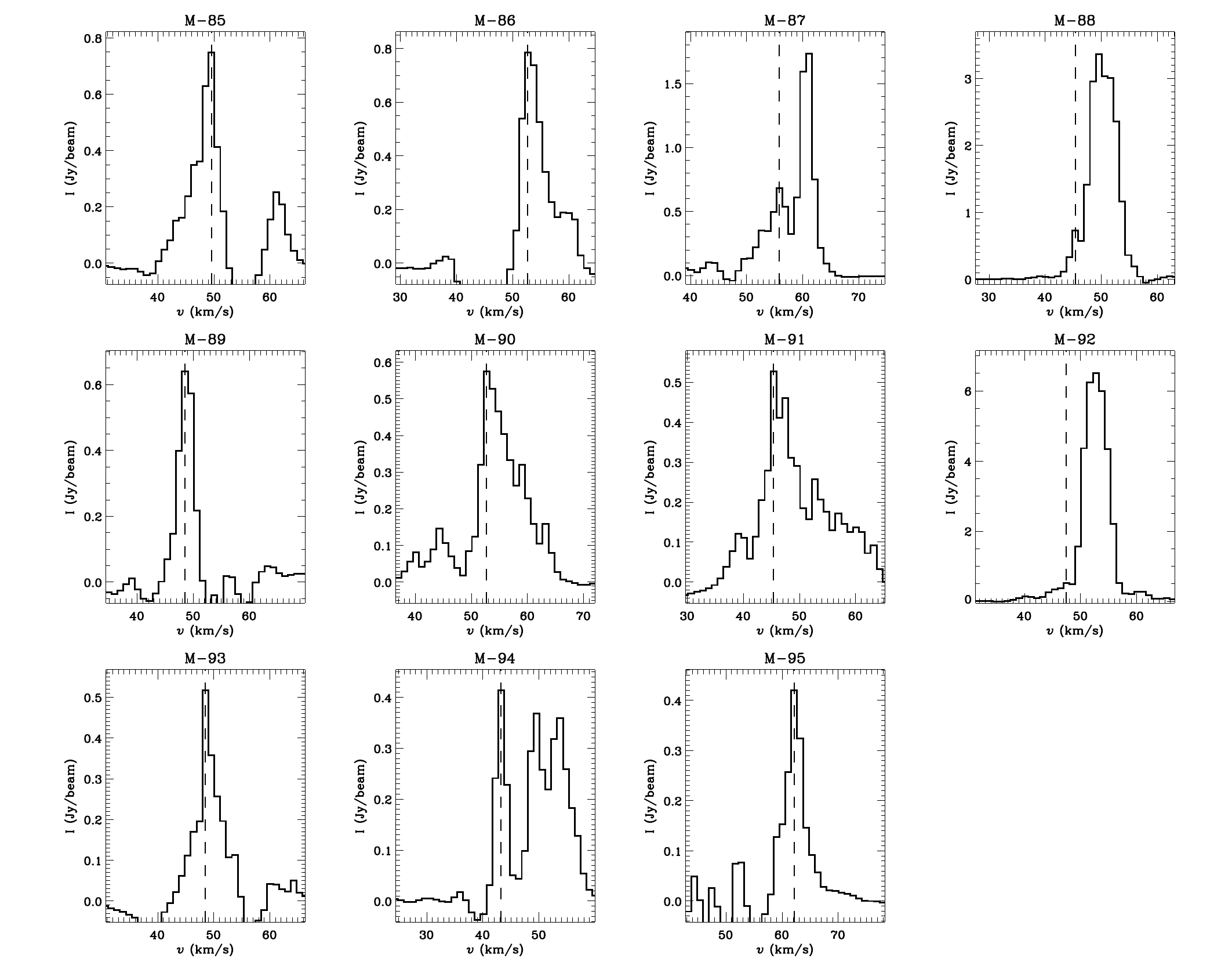} 
\caption{(Continued)}
\end{figure*}

\bibliographystyle{aasjournal}
\bibliography{M10.bib}

\end{document}